\documentclass[aps,prd,twocolumn,floatfix,superscriptaddress,groupedaddress,nofootinbib,preprintnumbers]{revtex4-2}
\usepackage[utf8]{inputenc}

\pdfoutput=1
\usepackage{graphicx}
\usepackage{latexsym} 
\usepackage{amsfonts}
\usepackage{amssymb}
\usepackage{amsmath}
\usepackage{xcolor}
\usepackage{cancel}
\usepackage{float}
\usepackage{multirow}
\usepackage[hidelinks]{hyperref}
\usepackage{physics}
\newcommand{\flowone}{\textsc{Flow-1}}
\newcommand{\flowtwo}{\textsc{Flow-2}}
\newcommand\geant{\textsc{Geant}4}
\newcommand{\scalo}{\textsc{SuperCalo}}
\newcommand{\icalo}{i\textsc{CaloFlow}}

\date{\today}
\begin{document}
\title{Calorimeter shower superresolution}

\author{Ian Pang}
\email{ian.pang@physics.rutgers.edu}
\affiliation{NHETC, Dept. of Physics and Astronomy, Rutgers University, Piscataway, NJ 08854, USA}

\author{John Andrew Raine}
\email{john.raine@unige.ch}
\affiliation{Département de physique nucléaire et corpusculaire, Université de Genève, 1211 Genève, Switzerland}

\author{David Shih}
\email{shih@physics.rutgers.edu}
\affiliation{NHETC, Dept. of Physics and Astronomy, Rutgers University, Piscataway, NJ 08854, USA}
\begin{abstract}
Calorimeter shower simulation is a major bottleneck in the Large Hadron Collider computational pipeline. There have been recent efforts to employ deep-generative surrogate models to overcome this challenge. However, many of best performing models have training and generation times that do not scale well to high-dimensional calorimeter showers. In this work, we introduce \scalo, a flow-based superresolution model, and demonstrate that high-dimensional fine-grained calorimeter showers can be quickly upsampled from coarse-grained showers. This novel approach presents a way to reduce computational cost, memory requirements and generation time associated with fast calorimeter simulation models. Additionally, we show that the showers upsampled by \scalo{} possess a high degree of variation. This allows a large number of high-dimensional calorimeter showers to be upsampled from much fewer coarse showers with high-fidelity, which results in additional reduction in generation time.
\end{abstract}
\maketitle
\section{Introduction}
The Large Hadron Collider (LHC) stands as a remarkable scientific instrument, providing invaluable insights into the fundamental building blocks of our universe. However, the LHC's ambitious exploration comes with computational challenges, particularly in the task of simulation (see \cite{Bruning:2015dfu, Calafiura:2729668, Software:2815292, Collaboration:2802918, HEPSoftwareFoundation:2020daq} for recent reviews of the current status and future plans for LHC computing). Simulating high-energy particle interactions in calorimeters poses the most significant computational bottleneck (see e.g., Fig.~1 of \cite{Calafiura:2729668}), necessitating the development of efficient and scalable simulation methods.

In recent years, deep learning-based surrogate models, such as Generative Adversarial Networks (GANs), Variational Autoencoders (VAEs), normalizing flows (NFs), and diffusion models~\cite{Paganini:2017hrr,Paganini_2018,deOliveira:2017rwa,Erdmann:2018kuh,Erdmann:2018jxd,Belayneh:2019vyx,Buhmann:2020pmy,ATL-SOFT-PUB-2020-006,Krause:2021ilc,Krause:2021wez,Buhmann:2021lxj,buhmann2021fast,Buhmann:2021caf,ATLAS:2021pzo,Mikuni:2022xry,ATLAS:2022jhk,Adelmann:2022ozp,Krause:2022jna,Cresswell:2022tof,AbhishekAbhishek:2022wby,schnakegenerating, Diefenbacher:2023vsw,Diefenbacher:2023prl,Buhmann:2023bwk,Hashemi:2023ruu,Dubinski:2023fsy,Acosta:2023zik,Amram:2023onf}, have shown promising results in accelerating the simulation process. 

To date, most of the existing surrogate modeling approaches have attempted to directly simulate the full-dimensionality calorimeter showers in a single step, i.e. transforming noise $z$ directly to calorimeter showers $x$.\footnote{Two exceptions are \cite{Diefenbacher:2023vsw,Buckley:2023daw}, which generated showers sequentially, layer by layer, conditioned on previous layers. This is quite different than the approach being proposed here, which will upsample coarse-grained showers all-at-once into higher-resolution showers.}
This direct, one-step approach presents serious challenges for generative models when $x$ is very high-dimensional. In this work we instead explore an alternative paradigm for the first time, which we dub \scalo, that applies superresolution techniques to high-dimensional calorimeter simulations. 

Superresolution methods based on deep learning have shown exceptional promise in image processing tasks, enhancing image quality and reconstructing high-resolution images from low-resolution inputs (see e.g.,~\cite{dong2015image,kim2016accurate,lim2017enhanced,ledig2017photo, zhang2018image,wang2018esrgan,Baldi:2020hjm,DiBello:2020bas,Erdmann:2023ngr}). However, many of these existing methods generate the high-resolution images in a deterministic way, whereas our problem requires a probabilistic reconstruction of high-dimensional calorimeter showers given lower resolution calorimeter showers. 

Following the bulk of the existing approaches to fast calorimeter simulation, we will focus on representing calorimeter showers in the form of 3-D images that are binned in position space.\footnote{Recently, there have been some works on generating much higher-resolution GEANT4 hits as point clouds instead of voxels~\cite{Buhmann:2023bwk, Acosta:2023zik,schnakegenerating}. We believe our \scalo\ framework could also be straightforwardly extended to these setups as well.} In this representation, the calorimeter shower geometry is made up of voxels (volumetric pixels).  In this work, we refer to these voxels interchangeably as fine voxels.
Then the ultimate goal of super-resolving calorimeter showers is to learn 
\begin{equation}
p(\vec E_{\rm fine}|\vec E_{\rm coarse}).
\end{equation}
Here, $\vec E_{\rm fine}$ are all the fine voxel energies and $\vec E_{\rm coarse}$ is a coarse-grained representation of them. 

Of course, if we tried to learn $p(\vec E_{\rm fine}|\vec E_{\rm coarse})$ with a single model, it would be no better in terms of model size than the original problem of learning $p(\vec E_{\rm fine}|E_{\rm inc})$. Hence, we are exploring physically-motivated approximations to the full density. 

One motivated ansatz is to say that each coarse voxel is upsampled to its fine voxels with a universal superresolution function that may be conditioned on some details such as the coarse voxel location and neighboring coarse voxel energies:
\begin{equation}
p(\vec E_{\rm fine}|\vec E_{\rm coarse}) = \prod_{i=1}^{N_{\rm coarse}}p(\vec e_{\text{fine},i}|E_{\text{coarse},i},\dots).
\end{equation}
Here $\vec e_{\text{fine},i}$ are the fine voxel energies associated with the $i$th coarse voxel. 
This ansatz assumes that after upsampling, $\vec e_{\text{fine},i}$ are all independent of each other. This adds importance to the choice of coarse representation, an issue we will explore in this work.

We test this method of generating fine-grained voxel energy showers by upsampling coarse-grained voxel energy showers on Dataset 2~\cite{CaloChallenge_ds2} of the \textit{Fast Calorimeter Simulation Challenge 2022} (\textit{CaloChallenge})~\cite{calochallenge}. The goal of the \textit{CaloChallenge} is to encourage the development of fast and high-fidelity calorimeter shower surrogate models. It provides three datasets of increasing dimensionality to serve as benchmarks for the community. In this work, we only focus on the second of the three datasets.

\begin{figure*}
    \centering
    \includegraphics[width=\textwidth]{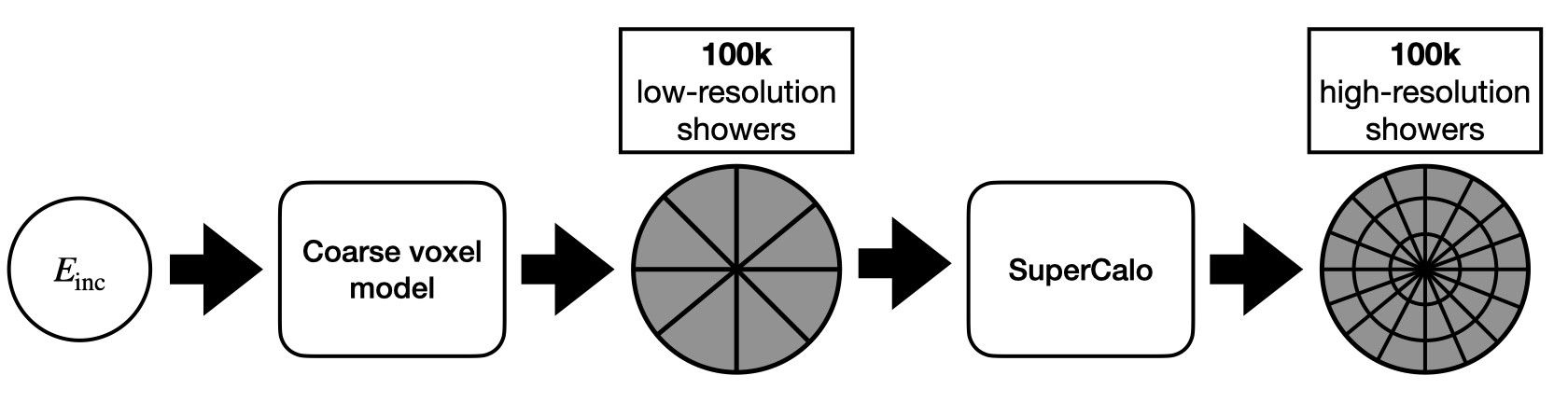}
    \caption{Schematic of full generation chain where the coarse voxel model (comprising of \flowone{}~and \flowtwo{}) and \scalo{}~are applied sequentially.}
    \label{fig:full_chain}
\end{figure*}

In Section~\ref{sec:dataset}, we describe the details of Dataset 2 of the \textit{CaloChallenge}. In Section~\ref{sec:supercalo}, we describe two possible choices of coarse voxelization and the resulting \scalo{} setup based on each choice. We also compare the performance of \scalo{} for the two choices of coarse voxelization. Next, we describe how we generate the coarse voxel energies with a flow-based setup in Section~\ref{sec:generate_coarse}. In Section~\ref{sec:results}, we use this setup together with \scalo{} to generate the fine voxel energies given the incident energies of the incoming particles. A schematic of this full model chain is shown in Fig.~\ref{fig:full_chain}. We evaluate the performance of the full model chain by comparing generated and reference distributions, by using classifier-based metrics (as in \cite{Krause:2021ilc,Krause:2021wez,Krause:2022jna,Diefenbacher:2023vsw}), and also by performing timing studies. Finally, we conclude in Section~\ref{sec:conclusion}.
\section{Dataset}
\label{sec:dataset}
\begin{figure}
    \centering
    \includegraphics[width=0.9\columnwidth,trim={0 3.4cm 0 3.4cm},clip]{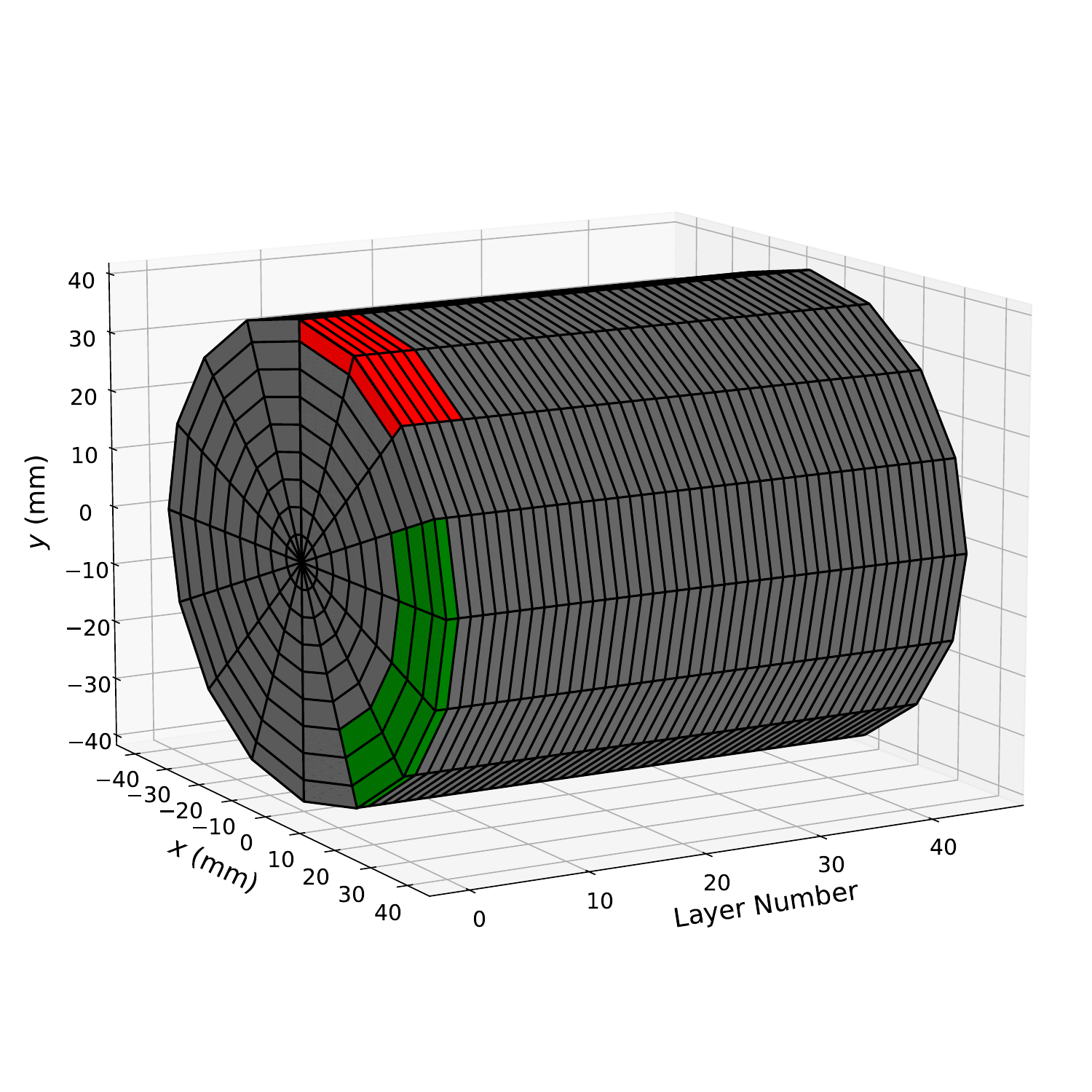}
    \caption{3-D calorimeter voxel geometry for Dataset 2 showing fine voxelization (gray) and representative coarse voxels for \scalo{}~$A$ (red) and \scalo{}~$B$ (green).}
    \label{fig:ds2_layer_geom}
\end{figure}

Dataset 2 consists of two sets of 100k electron showers generated using \geant{}~\cite{Agostinelli:2002hh,1610988,ALLISON2016186}. One set is used for training, while the second set is used for evaluation. For each shower, the dataset includes the incident energy $E_{\rm inc}$ of the incoming particle and 6480 fine voxel energies $\vec E_{\rm fine}$. The incident energy $E_{\rm inc}$ is log-uniformly distributed between 1~GeV and 1~TeV. The calorimeter geometry comprises 45 concentric cylindrical layers stacked along the direction of particle propagation ($z$); each layer is further divided into 16 angular bins ($\alpha$) and nine radial bins ($r$). In this work, we refer to these layers synonymously as fine layers. The geometry is identical for all the layers, and a diagram of the full 3-D calorimeter voxel geometry is shown in Fig.~\ref{fig:ds2_layer_geom}.

\section{\scalo}
\label{sec:supercalo}
In this section, we introduce \scalo{} and demonstrate that the fine voxel energies can be upsampled with high fidelity from the coarse voxel energies using a conditional NF. We outline the basics of NFs in Appendix~\ref{sec:NFs}. It is important to note that the general idea is not restricted to the use of NFs, and instead one can use alternative architectures in place of NFs. This makes \scalo{} an extremely flexible framework in the context of fast calorimeter simulation.

\subsection{Coarse voxelization}
\label{sec:coarse_vox}
The coarse voxel geometry is obtained from the full voxel geometry of the dataset by grouping neighboring fine voxels to form coarse voxels. There are many possible choices of coarse voxelization for each given full voxel geometry. In this work, we study two such choices and compare the fidelity of the upsampled showers against the \geant{} reference. These two coarse voxelizations are visualized in Fig.~\ref{fig:ds2_layer_geom}. We discuss the advantages and disadvantages of each choice of coarse voxelization and compare samples generated by the two approaches. 

\begin{description}

\item[\textbf{Choice A}] 1 coarse voxel = 1 $r$ $\times$ 2 $\alpha$  $\times$ 5 $z$.

Here, we group voxels in the $\alpha$ and $z$ directions (not in the $r$ direction). This choice of coarse voxelization results in a coarse voxel geometry with nine coarse layers, nine radial bins and eight angular bins which is comparable to the full voxel geometry found in the pion sample of Dataset 1~\cite{CaloChallenge_ds1}. Furthermore, coarse-graining in the $z$ direction allows the inter-layer correlations between fine voxels within each coarse voxel to be learned by the flow. The downside is that correlations among fine voxels within adjacent coarse voxels in the $r$ and $\alpha$ directions are not captured.

\item [\textbf{Choice B}] 1 coarse voxel = 3 $r$ $\times$ 4 $\alpha$ $\times$ 1 $z$.

Here, we group voxels in the $\alpha$ and $r$ directions (not in the $z$ direction). This choice of coarse voxelization results in a coarse voxel geometry with 45 layers, three radial bins and four angular bins. This choice of coarse voxelization helps to ensure that the total energy deposited within individual fine layers (layer energies) are learned correctly. However, it does not guarantee the correct inter-layer correlations between fine voxels. 

\end{description}

For each choice of coarse voxelization, we construct a superresolution flow to generate the relative\footnote{As in, normalized to sum to one.} fine voxel energies, $\hat e_{{\rm fine}, i}$, in a coarse voxel given the coarse voxel energy, $E_{\text{coarse}, i}$. We recover the fine voxel energies by multiplying them by the associated coarse voxel energy. We refer to the two superresolution flows as \scalo{}~$A$ and \scalo{}~$B$.

Note that choices A and B are examples of two-dimensional coarse graining. We also explored three-dimensional coarse grainings but found them to lead to poorer quality results. In particular, the generated samples tend to have the downsides of both choices A and B.

As the energy distributions within showers are highly structured, dependent on the incoming particle and the position in the calorimeter, \scalo{} uses additional conditional inputs to generate $\hat e_{{\rm fine}, i}$:

\begin{itemize}
    \item Incident energy of the incoming particle, $E_{\rm inc}$
    \item Deposited energy in coarse voxel $i$, $E_{\text{coarse},i}$
    \item Fine layer energies of layers spanned by coarse voxel $i$ (5 layers for \scalo{}~$A$ and 1 layer for \scalo{}~$B$)
    \item Deposited energy in neighboring coarse voxels in $\alpha$, $r$ and $z$ directions
    \item One-hot encoded coarse layer number 
    \item One-hot encoded coarse radial bin 
\end{itemize} 

Note that there is maximum of 6 neighboring coarse voxels for each coarse voxel. For coarse voxels with fewer than 6 adjacent coarse voxels, the missing neighboring coarse voxel energies are padded with zeros. Also, the fine layer energies are not obtainable from the coarse shower information for choice A. However, they are obtainable from our ``\flowone{}" (see Table~\ref{tab:conditional} and Section~\ref{sec:generate_coarse}) which generates the 45 fine layer energies as an initial step prior to applying \scalo. It costs very little extra to have ``\flowone{}" generate the fine layer energies instead of only the coarse ones. We decided to include the fine layer energies as conditional inputs as doing so enabled \scalo{}~$A$ to more accurately model the distribution of
fine voxel energies across the layers. In principle, it is possible to provide only the coarse layer energies as conditional inputs to \scalo, should fine layer energies not be available.
The dimensions of the conditional inputs and the outputs of \scalo{} are summarized in Table~\ref{tab:conditional}.

\subsection{Architecture and training}
\begin{table}
    \centering
    \begin{tabular}{|c|c|c|c|c|} \hline
         & Conditionals &\begin{tabular}{@{}c@{}}Dim of \\ conditional\end{tabular} & Output &\begin{tabular}{@{}c@{}}Dim of \\ output\end{tabular}  \\ \hline\hline
\flowone & $E_{\rm inc}$ & 1  & $\vec E_{\text{layer}}$ & 45\\ \hline 
\flowtwo & $E_{\rm inc}, \vec E^{(\rm coarse)}_{\text{layer}}$  & 10& $\vec E_{\text{coarse}}$ & 648\\ \hline
\scalo{}~$A$ & $E_{\rm inc},E_{\text{coarse}, i}, ... $  & 31&$\hat e_{\text{fine},i}$ & 10\\ \hline
\scalo{}~$B$ & $E_{\rm inc},E_{\text{coarse}, i}, ... $  & 57&$\hat e_{\text{fine},i}$ & 12\\ \hline
    \end{tabular}
    \caption{The conditional inputs for each flow, and the features whose probability distributions are the output of each flow. The full list of conditional inputs for \scalo{}~$A$ and \scalo{}~$B$ is included in Sec.~\ref{sec:coarse_vox}.}
    \label{tab:conditional}
\end{table}

All the NF models in this work are chosen to be Masked Autoregressive Flows (MAFs)~\cite{papamakarios2017masked} with compositions of Rational Quadratic Splines (RQS)~\cite{durkan2019neural} as the transformation function. The RQS transformations are parameterized using neural networks known as MADE blocks~\cite{germain2015made}. The details of the architecture of \scalo{}~$A$ and \scalo{}~$B$ are summarized in Table~\ref{tab:flow_architecture}. Moreover, the training of all the NF models in this work is optimized using independent \textsc{Adam} optimizers~\cite{kingma2014adam}.

\begin{table*}
\begin{center}
\begin{tabular}{|c|c|c|c|c|c|c|c|c|}
\hline
{\multirow{2}{*}{}}&dimension of & number of & \multicolumn{3}{c|}{layer sizes} & number of & RQS\\[-0.2ex]
Model &base distribution & MADE blocks & input & hidden & output & RQS bins & tail bound \\
\hline
\hline
\flowone{} & 45 & 8 & 256  & $1\times 256$ & 1035 & 8 & 14\\
\flowtwo{} & 648 &8 & 648  & $1\times 648$ & 14904 &  8 & 6\\
\scalo{}~$A$ & 10 &8 & 128  & $2\times 128$ & 230 & 8 & 14\\
\scalo{}~$B$ & 12 &8 & 128  & $2\times 128$ & 276 & 8& 14\\
\hline
\end{tabular}
\caption{Summary of architecture of the various flow models in this work. For the hidden layer sizes, the first number is the number of hidden layers in each MADE block and the second number is the number of nodes in each hidden layer (e.g., $2\times 128$ refers to 2 hidden layers per MADE block with 128 nodes per hidden layer).}
\label{tab:flow_architecture}
\end{center}
\end{table*}

For \scalo{}~$A$ and \scalo{}~$B$, we trained each of them using the mean log-likelihood of the data evaluated on the flow output for a total of 40 epochs with a batch size of 60k coarse voxels. We used 70\% of the training dataset for training and 30\% for model selection. We used the OneCycle~\cite{smith2019super} learning rate (LR) schedule with a base (initial) LR of $2\times10^{-5}$ and maximum LR of $1\times10^{-3}$. The LR increases for the first 18 epochs, decreases for the next 18 epochs, and then ends with an annihilation phase (4 epochs) which gradually decreases the LR by a factor of 10 below the base LR. The epoch with the lowest test loss is selected for subsequent sample generation.

The preprocessing used during training is detailed in Appendix \ref{sec:preprocessing}.

\subsection{Comparison with other approaches}

NFs have been employed successfully to generate the response of calorimeters for particle showers.
However, as the granularity of showers increases the computational requirements, both in time and memory requirement, scale with $\mathcal{O}\left(D^2\right)$, where $D$ is the dimensionality of the data.

To overcome this bottleneck, \cite{Diefenbacher:2023vsw,Buckley:2023daw} use autoregressive generation of the individual detector layers instead of one-shot generation of the full detector response.
This in turn reduces the complexity to scale linearly with the number of layers, and quadratically only with the dimensionality of the data in a single layer.
Our approach differs in that we still attempt to generate the full shower, and use a coarse representation in combination with \scalo{} to recover the granularity of the shower.
For very high granularity showers both approaches could be complementary to one another.

As we are applying \scalo{} only to a coarse representation, the initial showers can be produced with any generative model.
Diffusion models, such as those in~\cite{Mikuni:2022xry,Buhmann:2023bwk,Amram:2023onf} have demonstrated state-of-the-art fidelity in shower generation, but require many passes through the same network solve the ODE trajectory from noise to showers.
Distillation techniques have been employed to reduce the generation time of diffusion based models~\cite{Mikuni:2023dvk,Leigh:2023zle}, however combining the shower generation with \scalo{} to reduce the overall computation cost could present a viable alternative. A timing study for \scalo{} is shown in Section \ref{sec:timing}.

\subsection{Comparing coarse voxelization choices}
\label{sec:compare_coarse}
Here we discuss the performance of \scalo{} where the fine voxel energies for 100k showers are upsampled from the true coarse voxel energies from the training dataset. We compare the performance of \scalo{} for the two choices of coarse voxelization against the reference \geant{} samples.

\begin{figure*}[ht]
\includegraphics[width=0.5\columnwidth]{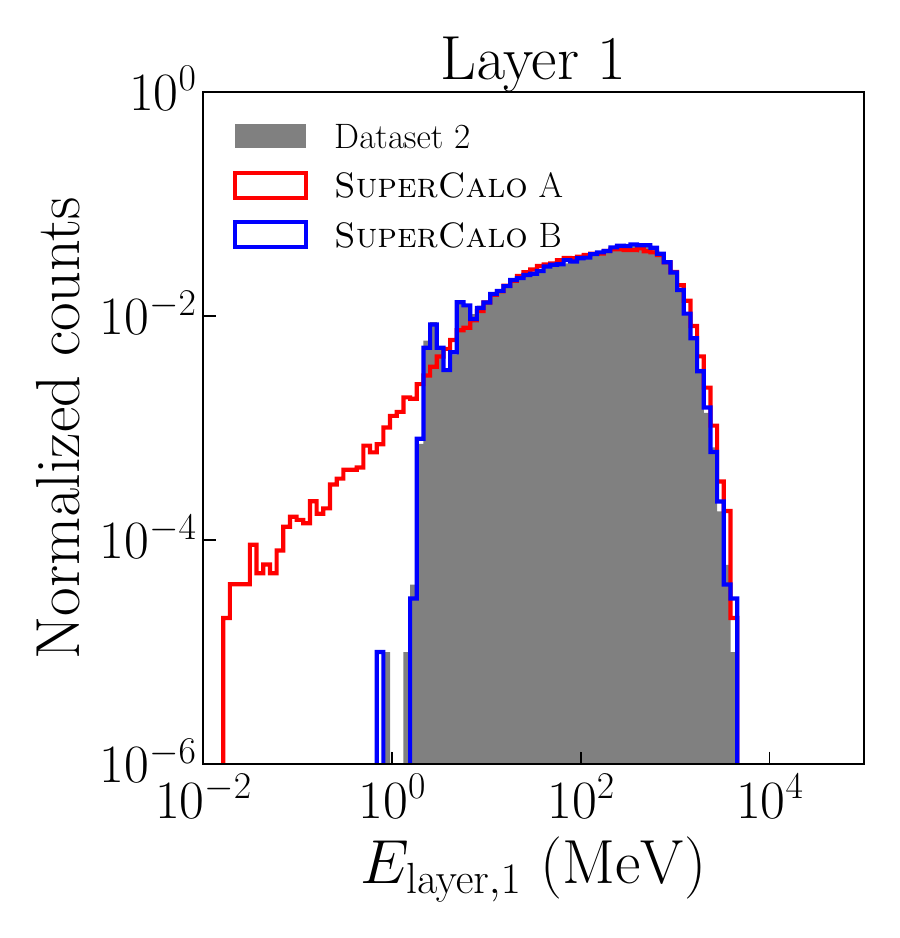}\includegraphics[width=0.5\columnwidth]{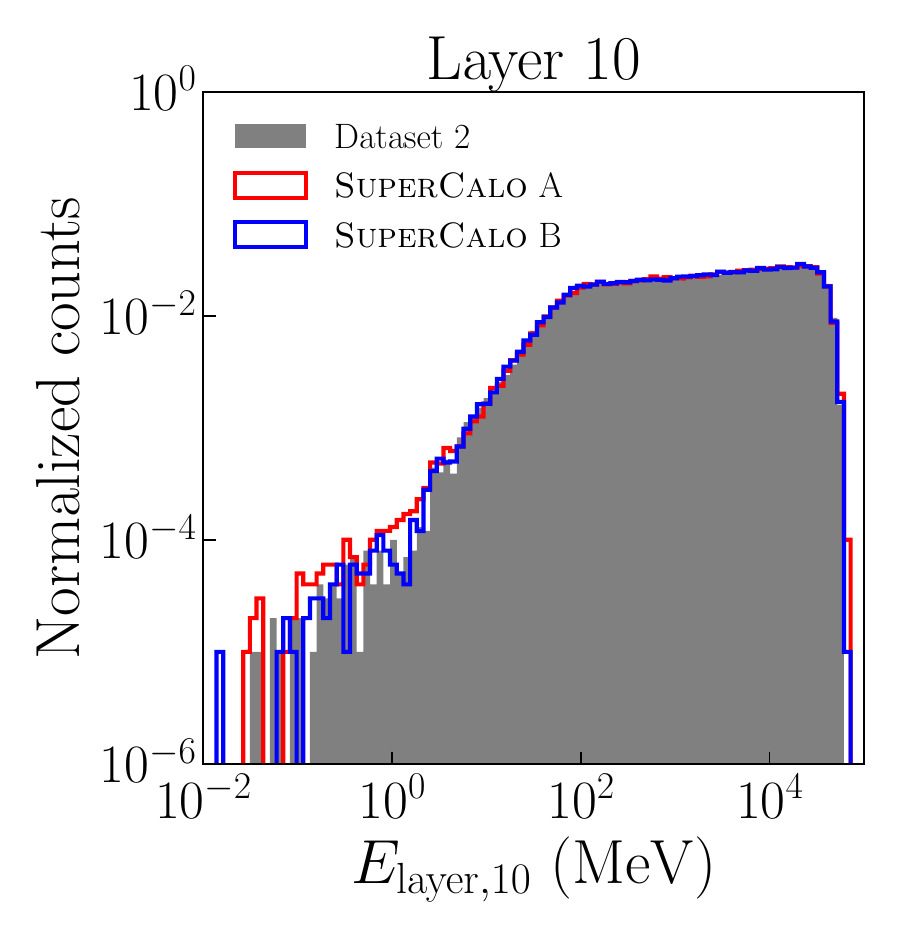}\includegraphics[width=0.5\columnwidth]{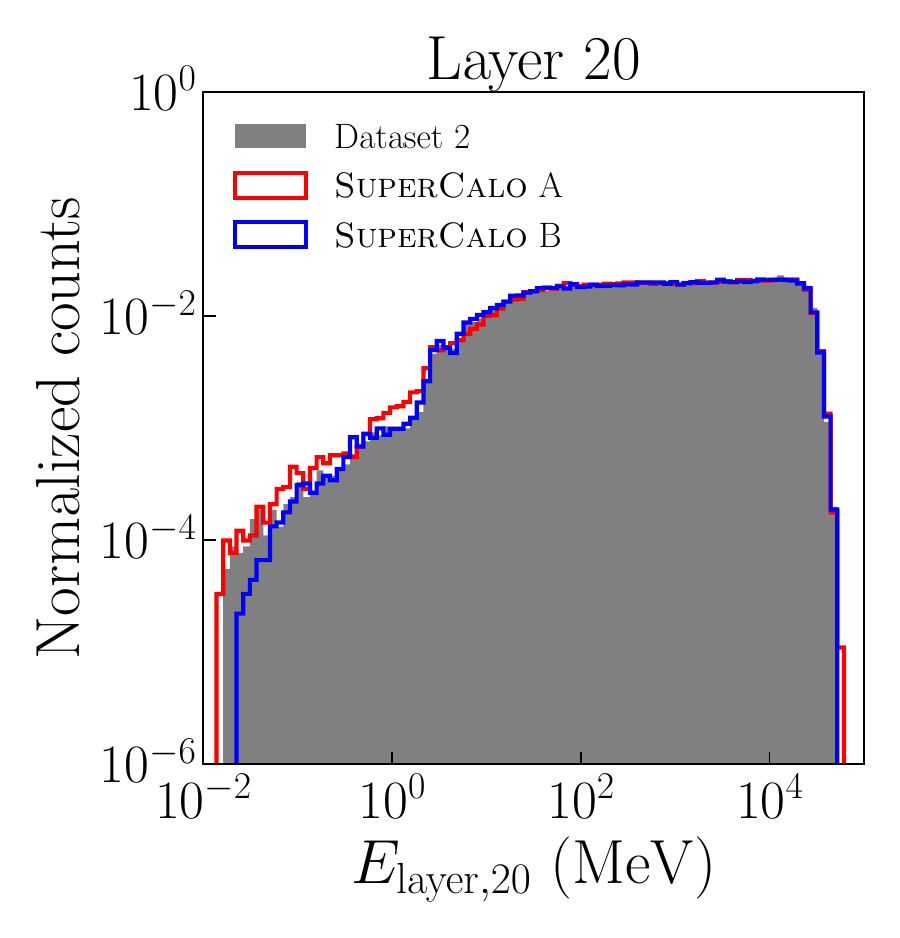}\includegraphics[width=0.5\columnwidth]{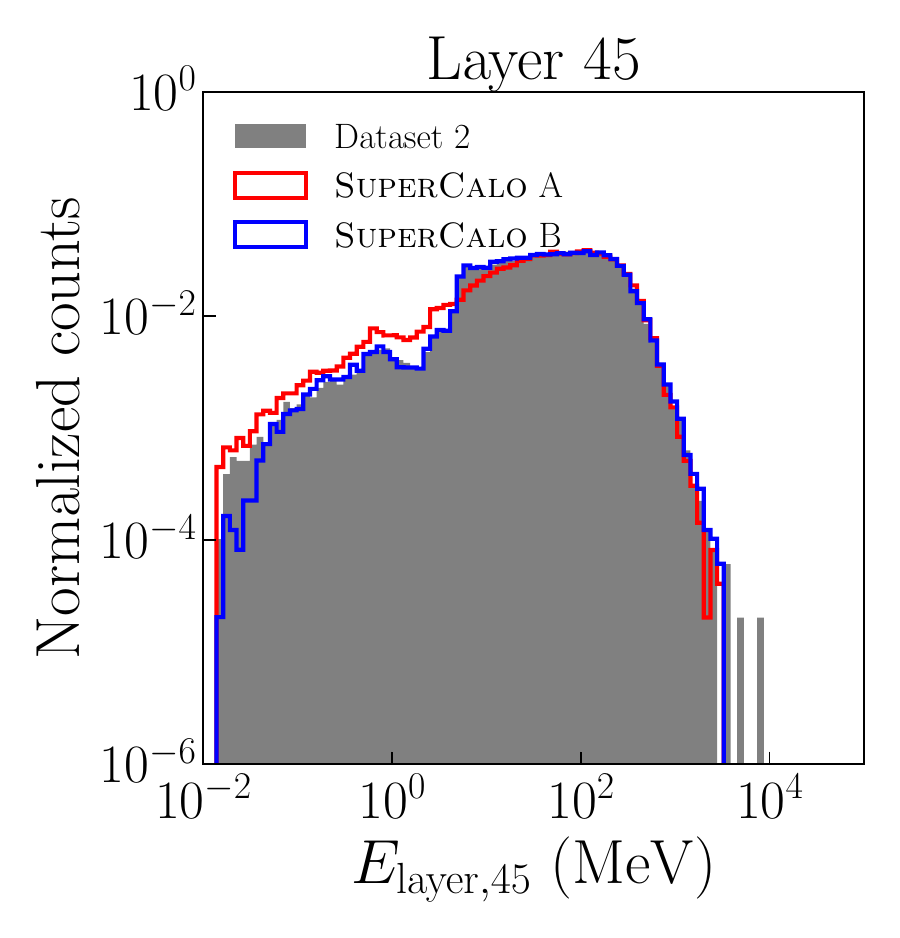}

\caption{Histograms of total energy deposited in a layer $i$ ($E_{{\rm layer},i}$), for $i=1$, 10, 20, and 45 (from left to right). Distribution of \geant\ data is shown in gray, and that of \scalo{}~$A$ ($B$) as red (blue) lines.}
\label{fig:layer_E_superres}
\end{figure*}

In Fig.~\ref{fig:layer_E_superres}, we show the distribution of layer energies in four selected layers (1st, 10th, 20th and 45th layers) to give a sense of the overall performance across all layers. It is clear from Fig.~\ref{fig:layer_E_superres} that \scalo{}~B (i.e.\ coarse-graining in $r$ and $\alpha$) results in better agreement in the layer energy distributions (especially for layer 1). This result is not surprising as each coarse voxel does not span multiple fine layers. By construction, we are guaranteed to model the layer energies $E_{\text{layer}, k}$ well as long as the generated fine voxels satisfy $\sum_{j=1}^{12} e_{\text{fine},ij} \approx E_{\rm coarse, i}$, where $i$ and $j$ are the coarse and fine voxel indices respectively. In contrast, each coarse voxel spans five fine layers for \scalo{}~A. Hence, for each coarse voxel, \scalo{}~A has to learn the correct distribution of fine voxel energies across five layers for the layer energies to be modelled well.
Furthermore, as all coarse voxels are considered independently, it is not possible for \scalo{}~A to capture the correlation between energy depositions across fine layers between multiple upsampled coarse voxels.

We also looked at histograms for other distributions, such as fine voxel energies and the center of energy in each layer. There is generally good agreement with the reference distribution with similar performance for both choices of coarse-graining. The generated distributions are similar to results that we show in Section~\ref{sec:results} and hence are not repeated here.

\renewcommand{\arraystretch}{1.5}
\begin{table}[!ht]
\begin{center}
\begin{tabular}{|c||c|c|c|c|}
\hline
Coarse& \multicolumn{2}{|c|}{low-level features} &  \multicolumn{2}{c|}{high-level features}\\
voxelization & AUC & JSD & AUC & JSD \\
\hline \hline
A ($z$ and $\alpha$) & \; 0.660(5) \; & \; 0.057(3) \; & \; 0.661(6) \; & \; 0.059(7) \; \\
\hline
B ($\alpha$ and $r$) & \;0.781(3) \; & \;0.186(4)  \; & \; 0.587(3) \; & \; 0.019(2) \; \\
\hline
\end{tabular}
\caption{Mean and standard deviation of 10 independent classifier runs trained on generated samples from \scalo{} with two different choices (A or B) of coarse voxelization versus \geant~samples. Note that the generated samples are obtained by upsampling the true coarse voxel energies from the training
dataset using \scalo{}.}
\label{tab:cls_results_upsample}
\end{center}
\end{table}
\renewcommand{\arraystretch}{1}

To fully capture any mismodelling of the correlations between features that are not reflected in 1-D histograms, we train a DNN binary classifier to distinguish between our generated showers and the reference samples (see Appendix \ref{sec:classifier_arch} for the classifier architecture and training procedure).

We train the classifier on both low-level features (LLFs) and high-level features (HLFs). LLFs refers to all fine voxel energies (normalized with $E_\text{inc}$ and multiplied by a factor 100) and $E_\text{inc}$ itself (preprocessed as $\log_{10} E_\text{inc}$). HLFs are the incident energy (preprocessed as $\log_{10}{E_{\text{inc}}}$), the fine layer energies (preprocessed as $\log_{10}{(E_{{\rm layer},i}+10^{-8})}$), the center of energy in the $x$ and $y$ directions (normalized with a factor 100), and the widths of the $x$ and $y$ distributions (normalized with a factor 100).

The classifier scores are summarized in Table \ref{tab:cls_results_upsample} using the area under the receiver operating characteristic curve (AUC). According to the Neyman-Pearson lemma, we expect the AUC to be 0.5 if the reference and generated probability densities are equal. The AUC is 1 if the classifier is able to perfectly distinguish between reference and generated samples. The second metric, JSD $\in [0, 1]$, is the Jensen-Shannon divergence which also measures the similarity between two binned probability distributions. The JSD is 0 if the two distributions are identical and 1 if they are disjoint. From Table \ref{tab:cls_results_upsample}, we see that the classifier scores are generally much less than unity which indicates that the generated samples are of sufficiently high fidelity to fool the classifier.

\begin{figure}
    \centering
    \includegraphics[width=0.7\columnwidth]{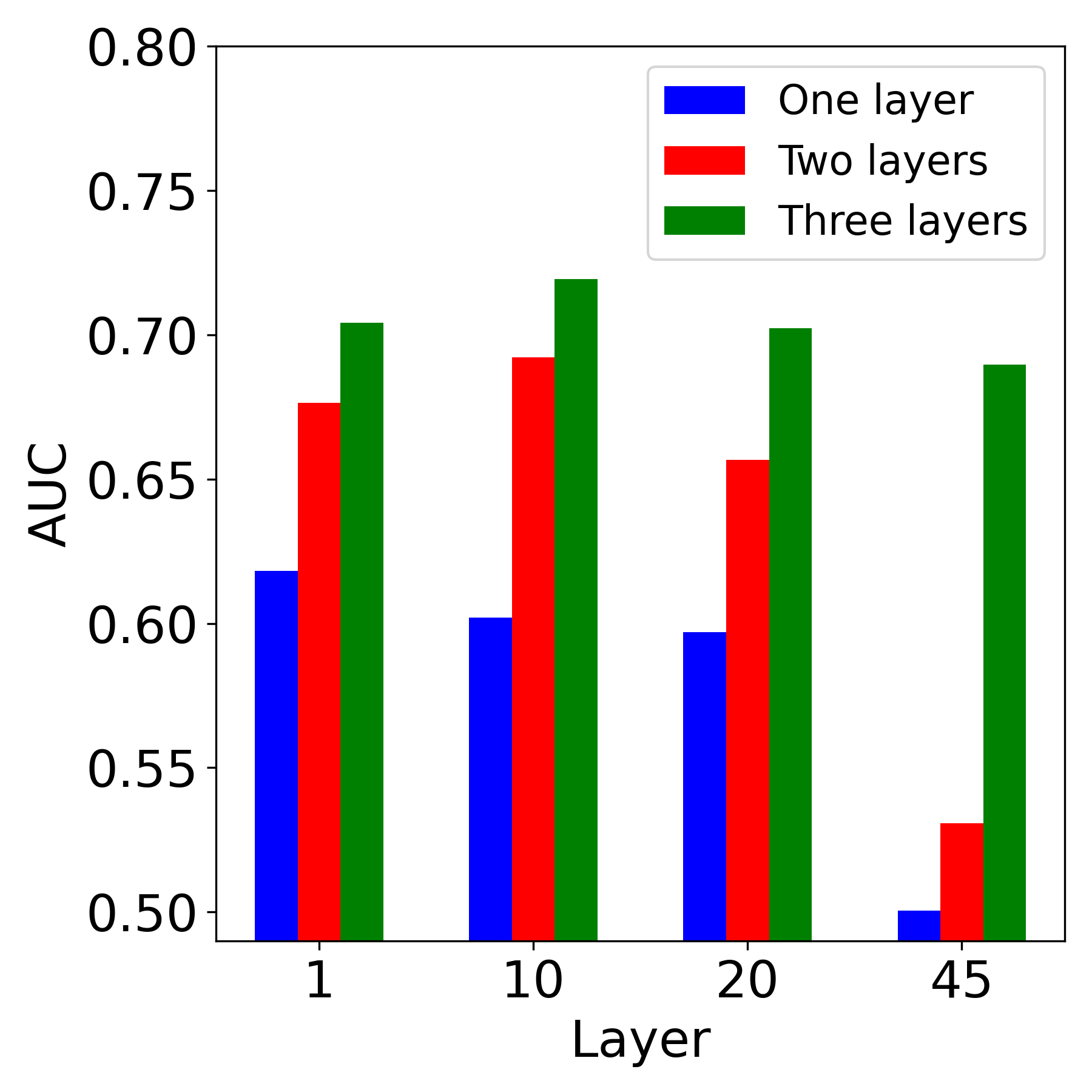}
    \caption{Plot of AUC scores for classifier trained on voxel energies from single layer (blue), two adjacent layers (red) and three adjacent layers (green). The horizontal axis indicates the layer number of the first layer's voxels that the classifier was trained on.
    For the final layer (45) the preceding adjacent layers are taken, otherwise subsequent adjacent layers are taken.}
    \label{fig:multilayer_classifier}
\end{figure}

We note that \scalo{}~$A$ has AUC $\sim$ 0.66 for both LLF and HLF classifiers. In contrast, \scalo{}~$B$ has a lower AUC of 0.587 for the HLF classifier and higher AUC of 0.781 for the LLF classifier. The low HLF classifier AUC of \scalo{}~$B$ is consistent with the layer energy histograms in Fig \ref{fig:layer_E_superres}. The high LLF classifier AUC of \scalo{}~$B$ can be explained by having incorrect inter-layer fine voxel energy correlations due to the choice of coarse voxelization. We tested this explanation by training a classifier just on voxel energies from a single layer from the reference and generated samples, and obtained a significantly lower AUC. Next, we trained a classifier on voxel energies from two adjacent layers, and this resulted in a higher AUC. We found that training a classifier on voxel energies from a larger number of adjacent layers resulted in higher AUCs. We observed this similar behavior even after repeating this experiment for adjacent layers belonging to the early, middle and later layers. See Fig.~\ref{fig:multilayer_classifier} for the illustration of this.

To visualize the failure of \scalo{}~$B$ in capturing inter-layer correlations, we regrouped the generated fine voxels from \scalo{}~$A$ and \scalo{}~$B$ according to coarse voxelization B. Next, we define the feature
\begin{equation}
\rho \equiv \frac{\vec{e}_{\rm fine}^{\ (n)} \cdot \vec{e}_{\rm fine}^{\ (n+1)}}{\abs{\vec{e}_{\rm fine}^{\ (n)}} \abs{\vec{e}_{\rm fine}^{\ (n+1)}}},
\end{equation}
which is the inner product of relative fine voxel energies associated with a pair of neighboring coarse voxels in the $z$-direction. Here the fine voxel energies in the coarse voxel are represented as a 12-dimensional vector $\vec{e}_{\rm fine}^{\ (n)}$, where $n$ is the index of the coarse voxel layer. 

\begin{figure*}[ht]
\includegraphics[width=0.67\columnwidth]{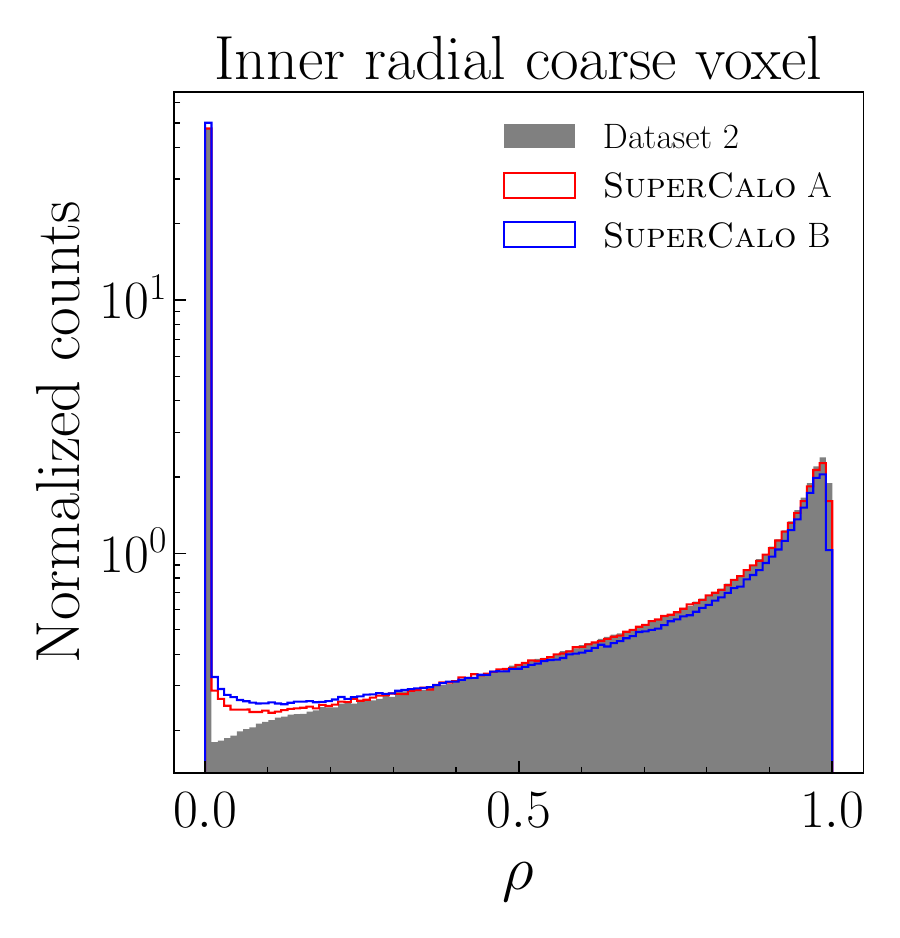}\includegraphics[width=0.67\columnwidth]{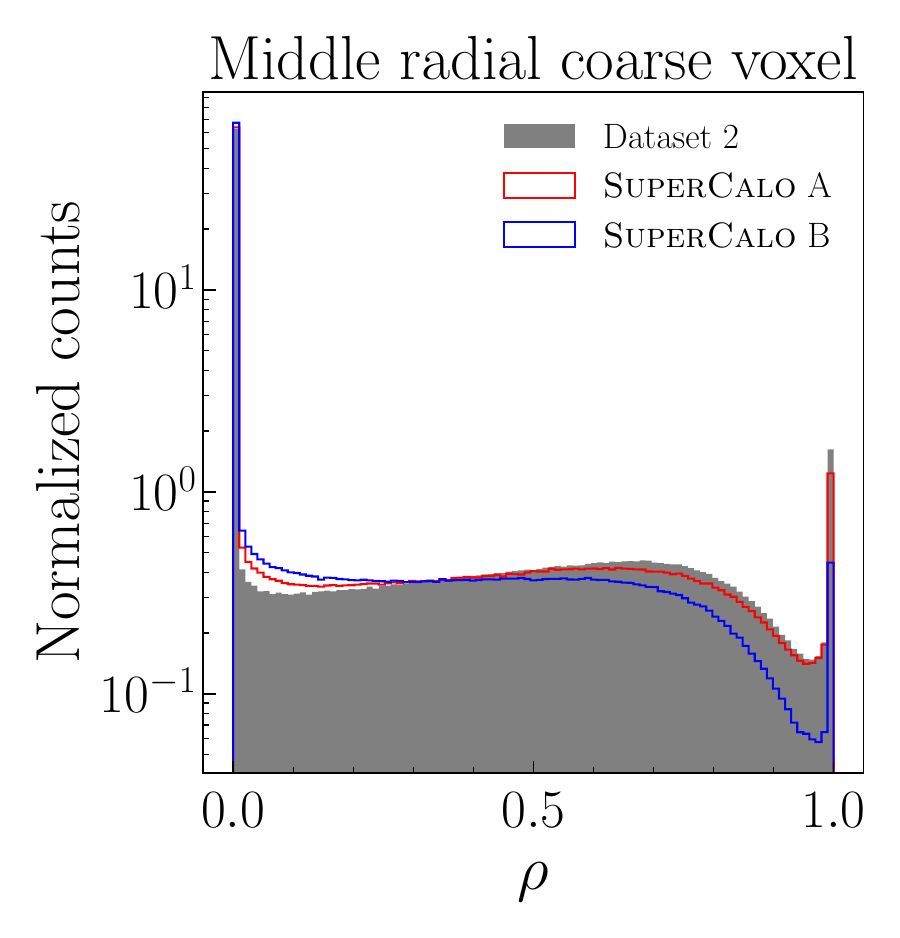}\includegraphics[width=0.67\columnwidth]{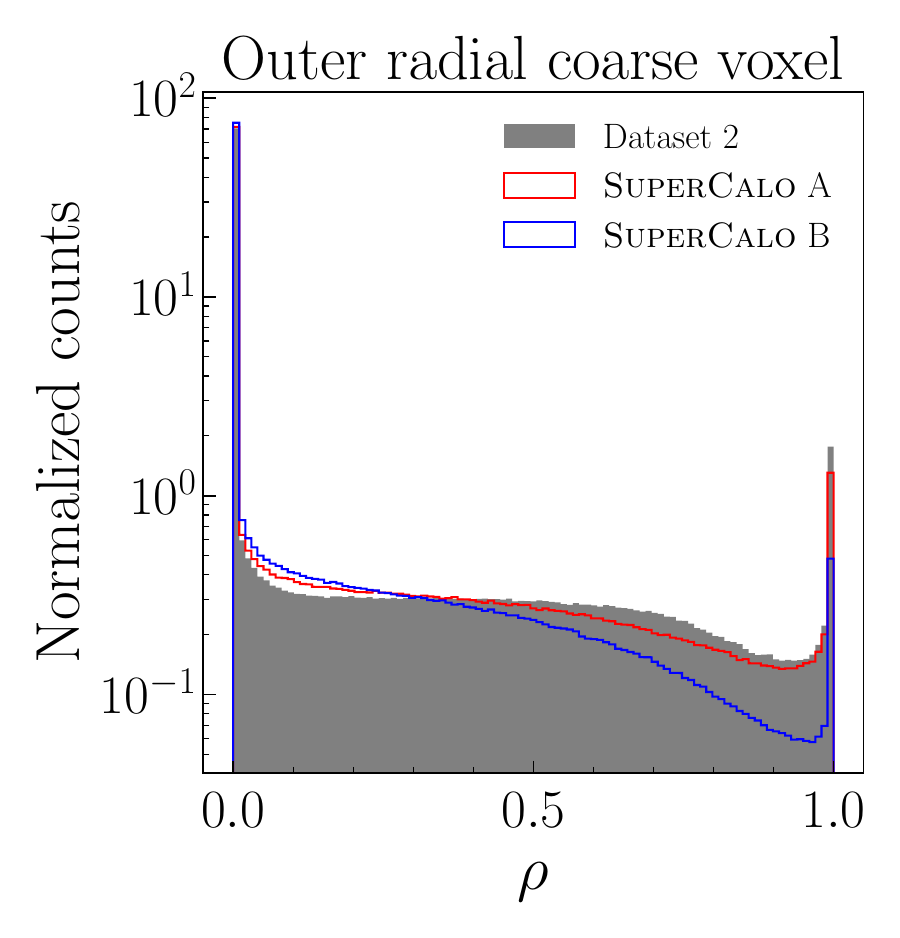}

\caption{Histograms of the $\rho$ distribution for inner, middle and outer coarse radial bins (from left to right). Distribution of \geant\ data is shown in gray, and that of \scalo{}~$A$ ($B$) as red (blue) lines. See text for definition of $\rho$.}
\label{fig:correlation_plot}
\end{figure*}

In Fig.~\ref{fig:correlation_plot}, we plot the distributions of $\rho$, marginalized over all adjacent layers ($n$, $n+1$), and broken down into the three coarse radial bins of voxelization B. We see that the generated $\rho$ distribution from \scalo{}~$B$ has substantial deviation from the reference distribution for coarse voxels in the middle and outer coarse radial bins. In contrast, the distribution from \scalo{}~$A$ agrees relatively well with the reference in all 3 coarse radial bins which is consistent with our observation that \scalo{}~A is learning the inter-layer fine voxel energy correlation better than \scalo{}~B.

\section{Generating coarse voxel energies with flows}

\label{sec:generate_coarse}

The goal of the \textit{CaloChallenge} is to design generative models to learn the full joint 
 probability density $p(\vec E_{\rm fine}|E_{\rm inc})$. To this end, we use two conditional NFs which we refer to as \flowone{} and \flowtwo{} as in \cite{Krause:2021ilc,Krause:2021wez, Krause:2022jna}. \flowone{} learns the distribution of total deposited energy in each fine layer conditioned on the incident energy of the particle $p_1(\vec E_{\rm layer}|E_{\rm inc})$. This flow is the same one used in \cite{Buckley:2023daw}. This allows us to generate the conditional inputs required for \flowtwo{} and \scalo{}. \flowtwo{} has the goal of learning the distribution of coarse voxel energies $p_2(\vec E_{\rm coarse}|E_{\rm inc},\vec E^{(\rm coarse)}_{\rm layer})$ conditioned on the incident energy of the particle and coarse layer energies $\vec E^{(\rm coarse)}_{\rm layer}$. Sampling sequentially with \flowone{}, \flowtwo{} and \scalo{} allows us to generate samples that approximately follow the distribution $p(\vec E_{\rm fine}|E_{\rm inc})$.

 As it is important for the model to capture the inter-layer correlations between fine voxels, we choose \scalo{}~A as the superresolution flow in the full model chain.\footnote{We experimented with combining samples from \scalo{}~A and \scalo{}~B. In particular, the fine voxels from layers 1-5 were generated using \scalo{}~$B$, while the other voxels were generated using \scalo{}~$A$. While this improved the layer energy $E_{{\rm layer},i}$ distributions for $i=1,2,..., 5$, the classifier scores did not improve significantly.}  The architectures of \flowone{} and \flowtwo{} are summarized in Table~\ref{tab:flow_architecture}. The details of the conditional inputs and the outputs of \flowone{} and \flowtwo{} are shown in Table~\ref{tab:conditional}, while the preprocessing used during training is detailed in Appendix \ref{sec:preprocessing}.

 Both \flowone{} and \flowtwo{} are MAFs with RQS transformations and are trained using the mean log-likelihood of the data evaluated on the output of the flows. We train with a batch size of 1000 for both flows. We used 70\% of the training dataset for training and 30\% for model selection. For \flowone{}, we trained for 500 epochs using the OneCycle LR schedule~\cite{smith2019super} with base LR of $1\times10^{-5}$ and maximum LR of $1\times10^{-4}$, which finishes with an annihilation phase that decreases the LR by a factor of 10 lower than the base LR. For \flowtwo{}, we trained for 40 epochs using the cyclic LR schedule~\cite{smith2017cyclical} with a cycle length of 10 epochs. The base and initial maximum LRs were chosen to be $5\times10^{-5}$ and $2\times10^{-3}$ respectively. The maximum LR is decreased by 20\% after each cycle. The epoch with the lowest test loss is selected for subsequent sample generation.

\section{Full chain results}
\label{sec:results}

As illustrated in Fig.~\ref{fig:full_chain}, we generate 100k electron showers using \flowone{}, \flowtwo{} and \scalo{}. The fidelity of the generated showers is determined by comparing against the reference \geant{} samples. As in Sec.~\ref{sec:compare_coarse}, we present the same four layers when comparing layer-level distributions to give a sense of the overall performance across the full depth of the shower.  
\subsection{Distributions}
\begin{figure*}[ht]
\includegraphics[width=0.5\columnwidth]{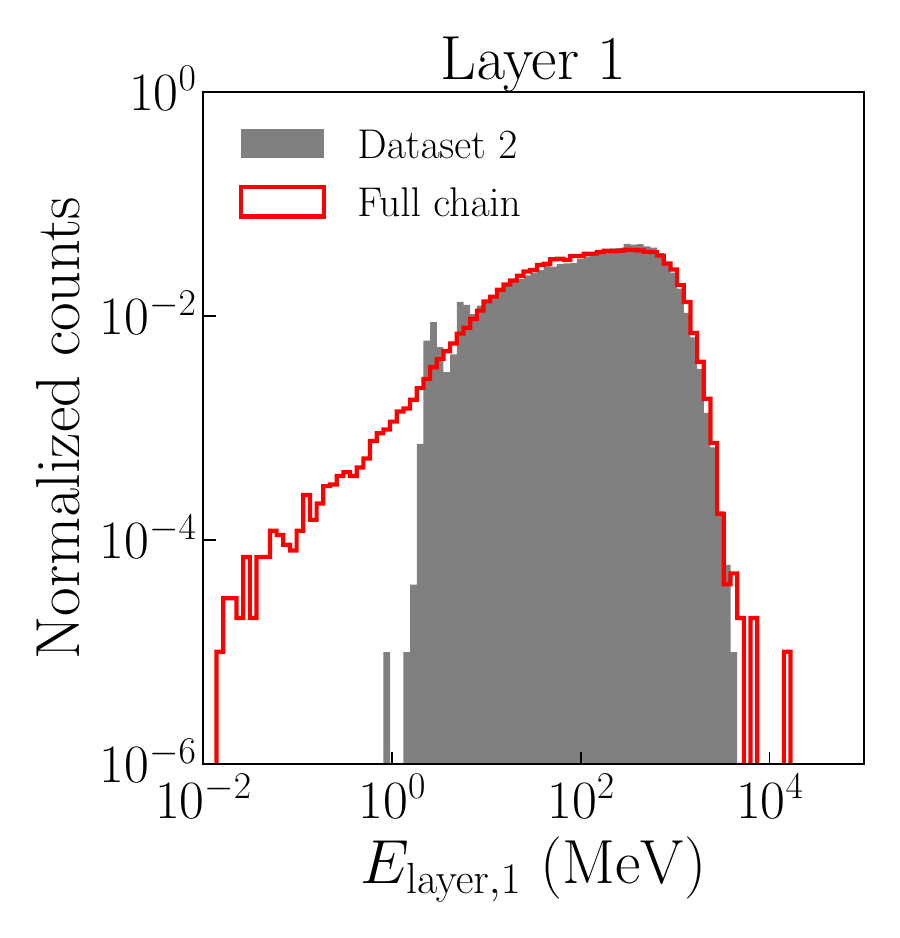}\includegraphics[width=0.5\columnwidth]{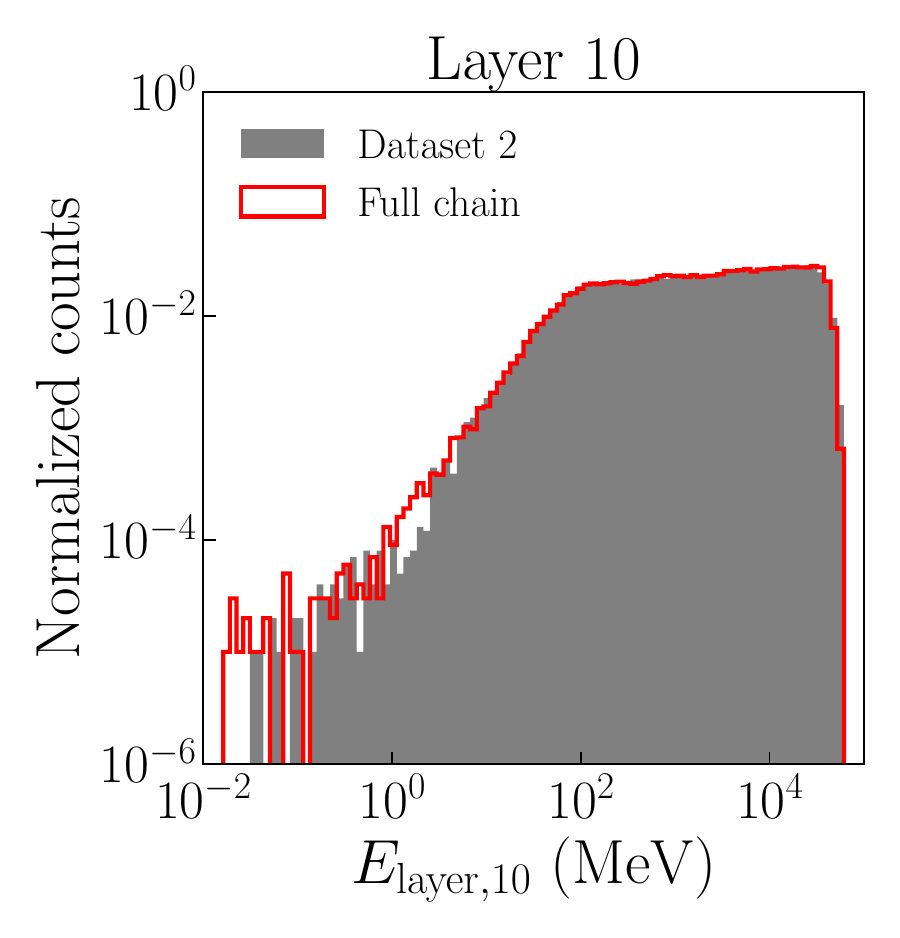}\includegraphics[width=0.5\columnwidth]{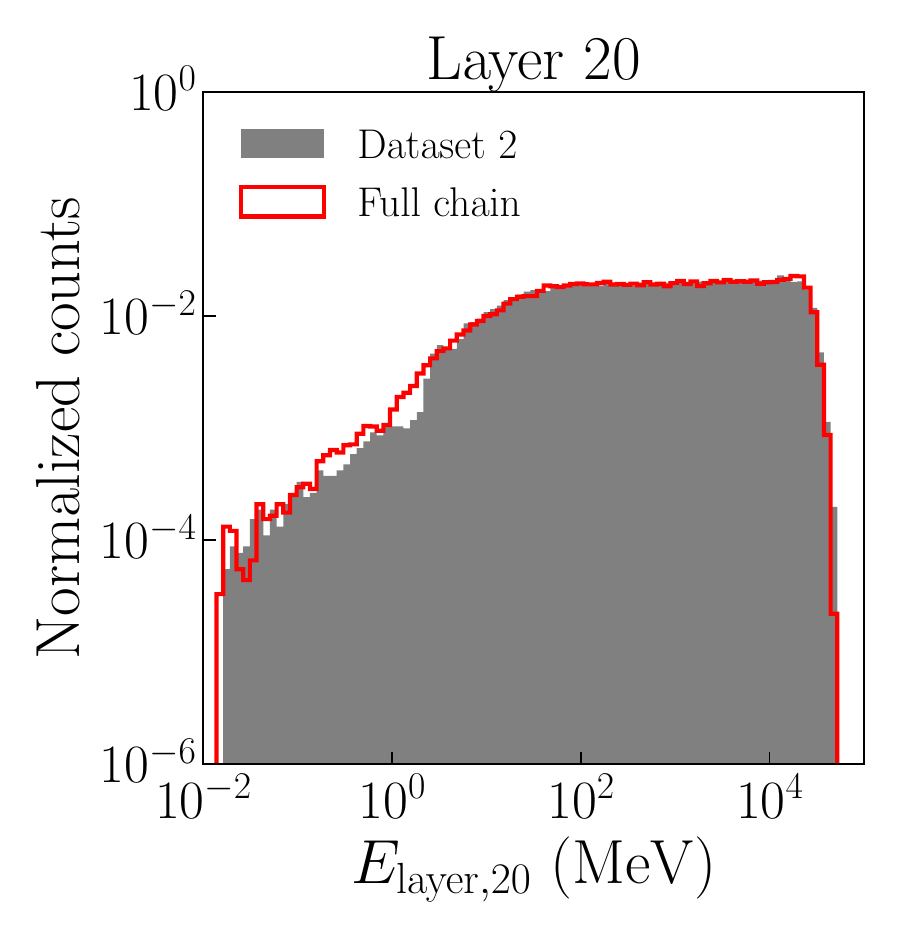}\includegraphics[width=0.5\columnwidth]{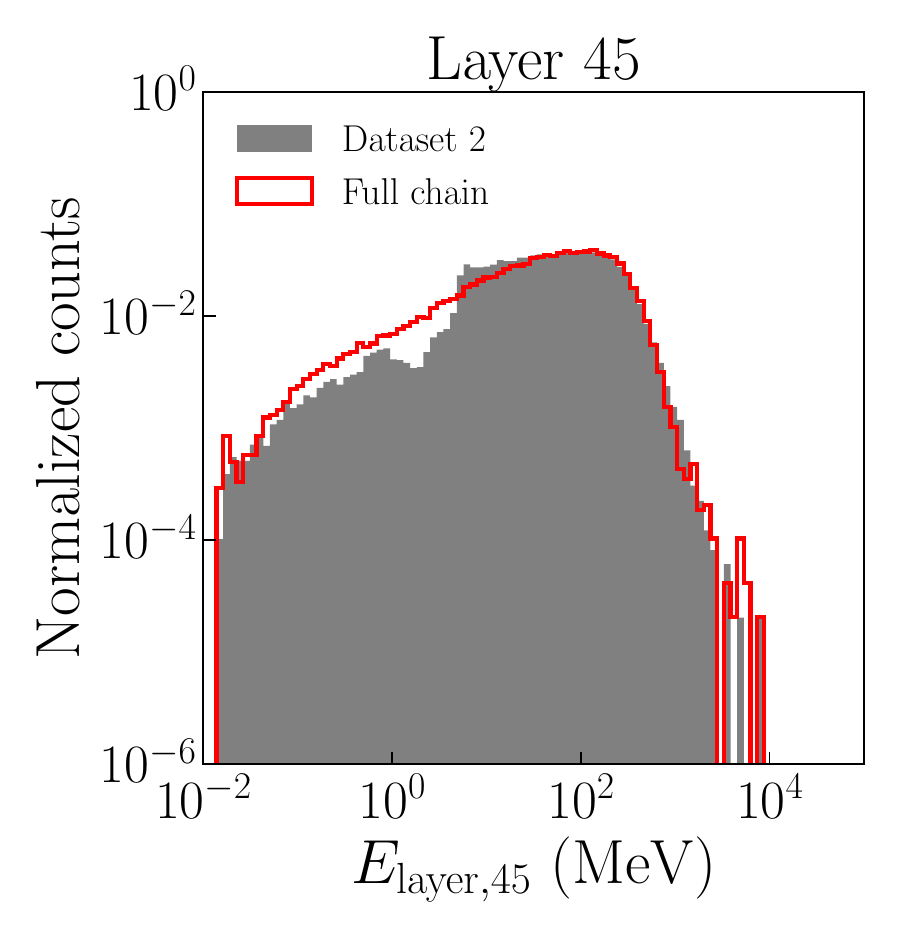}

\caption{Histograms of total energy deposited in a layer $i$ ($E_{{\rm layer},i}$), for $i=1$, 10, 20, and 45 (from left to right). Distribution of \geant\ data is shown in gray, and that of our full model chain as red lines.}
\label{fig:layer_E}
\end{figure*}

In Fig.~\ref{fig:layer_E}, we can see the total deposited energy in each layer. There is relatively good agreement with the reference distribution for the most of the layers with the exception of layer 1. For layer 1, we observe that the generated layer energy distribution extends much lower than the minimum layer energy of the reference distribution. This mismatch of layer 1 energies was already seen in Fig.~\ref{fig:layer_E_superres}, so it is a consequence of using \scalo{}~$A$ and not the fault of the coarse shower generator \flowtwo.

\begin{figure*}[ht]
\includegraphics[width=0.5\columnwidth]{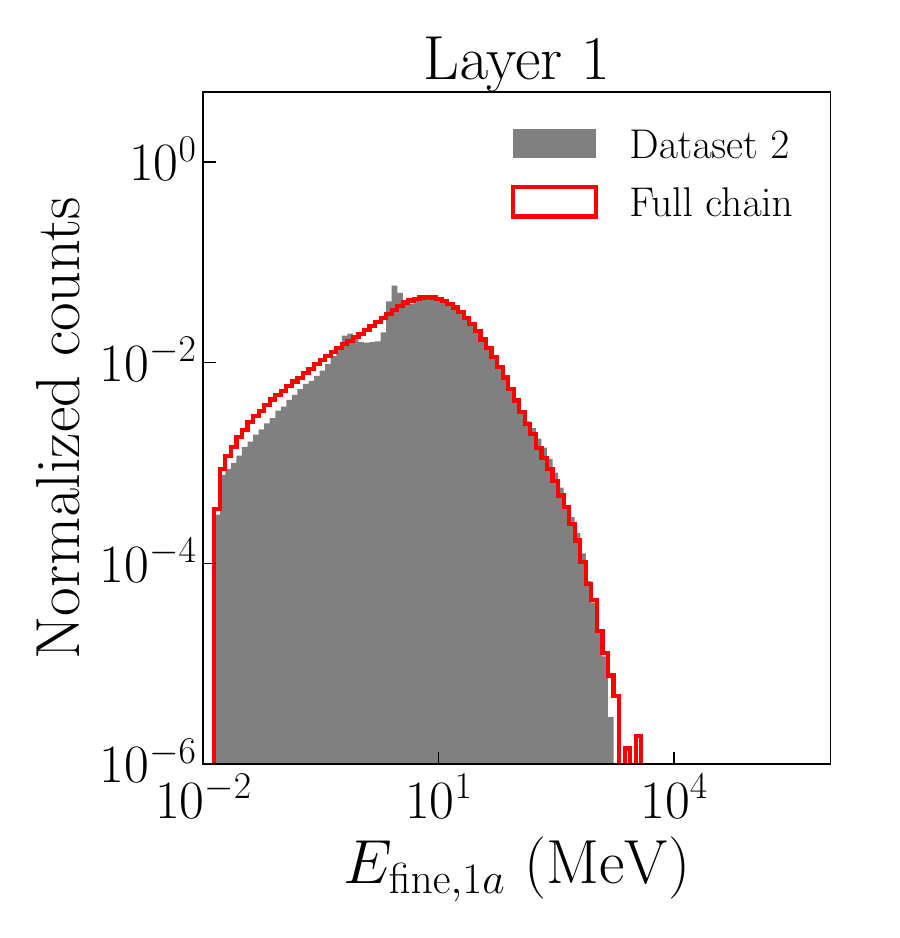}\includegraphics[width=0.5\columnwidth]{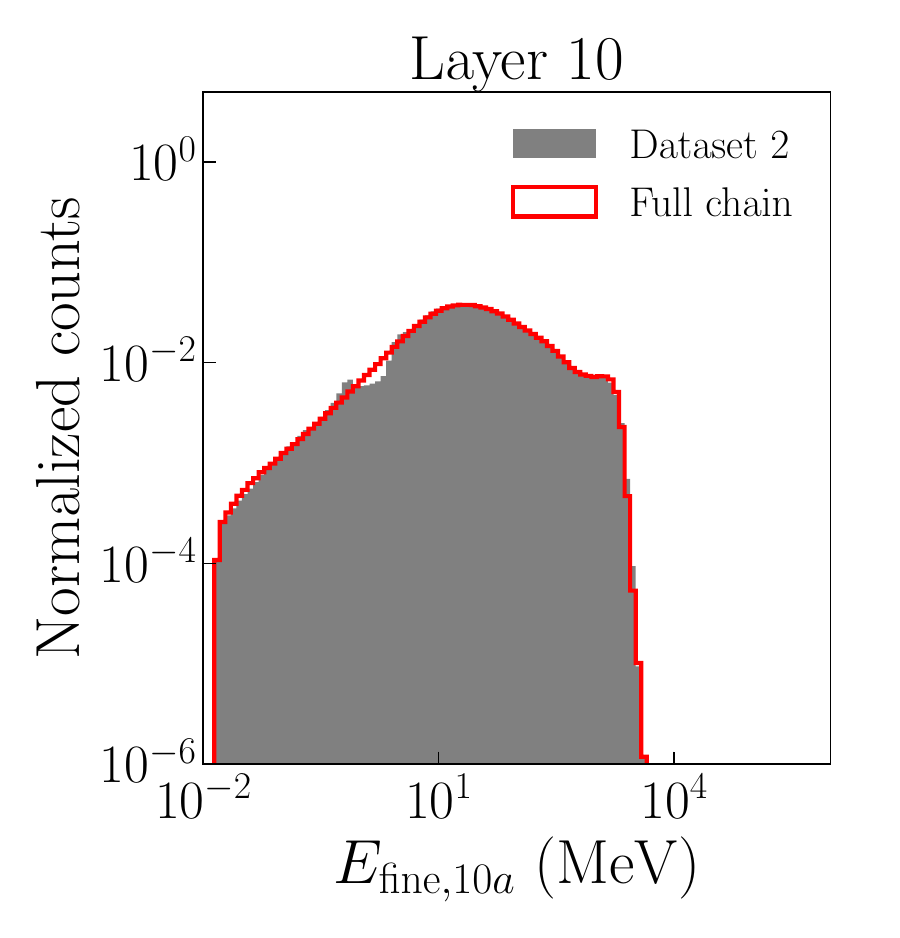}\includegraphics[width=0.5\columnwidth]{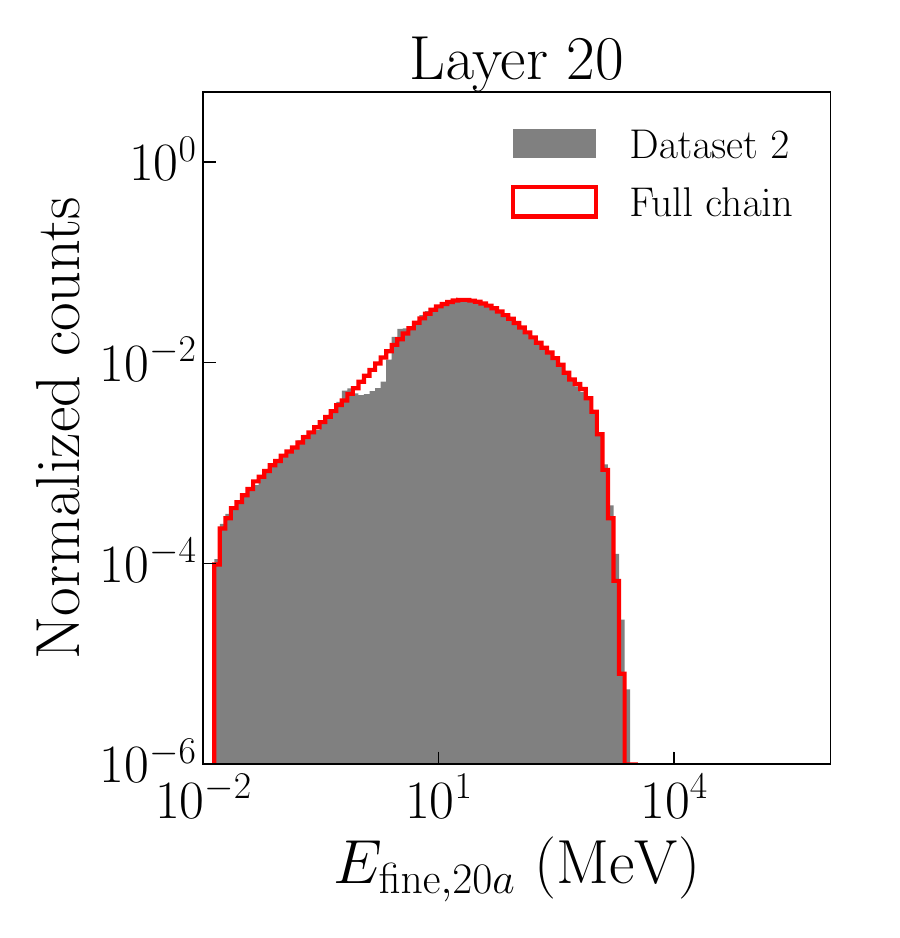}\includegraphics[width=0.5\columnwidth]{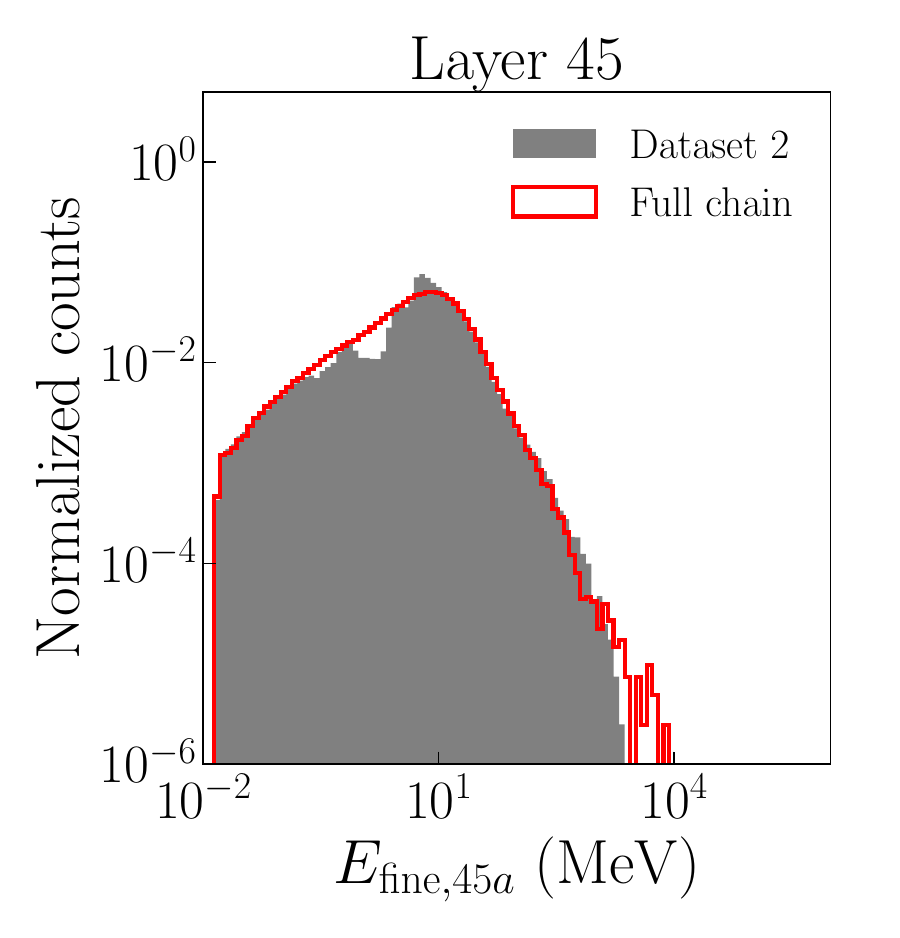}

\caption{Histograms of fine voxel energy distribution in a layer $i$, for $i=1$, 10, 20, and 45 (from left to right). Distribution of \geant\ data is shown in gray, and that of our full model chain as red lines.}
\label{fig:voxel_E}
\end{figure*}

\begin{figure}[ht]
\includegraphics[width=0.7\columnwidth]{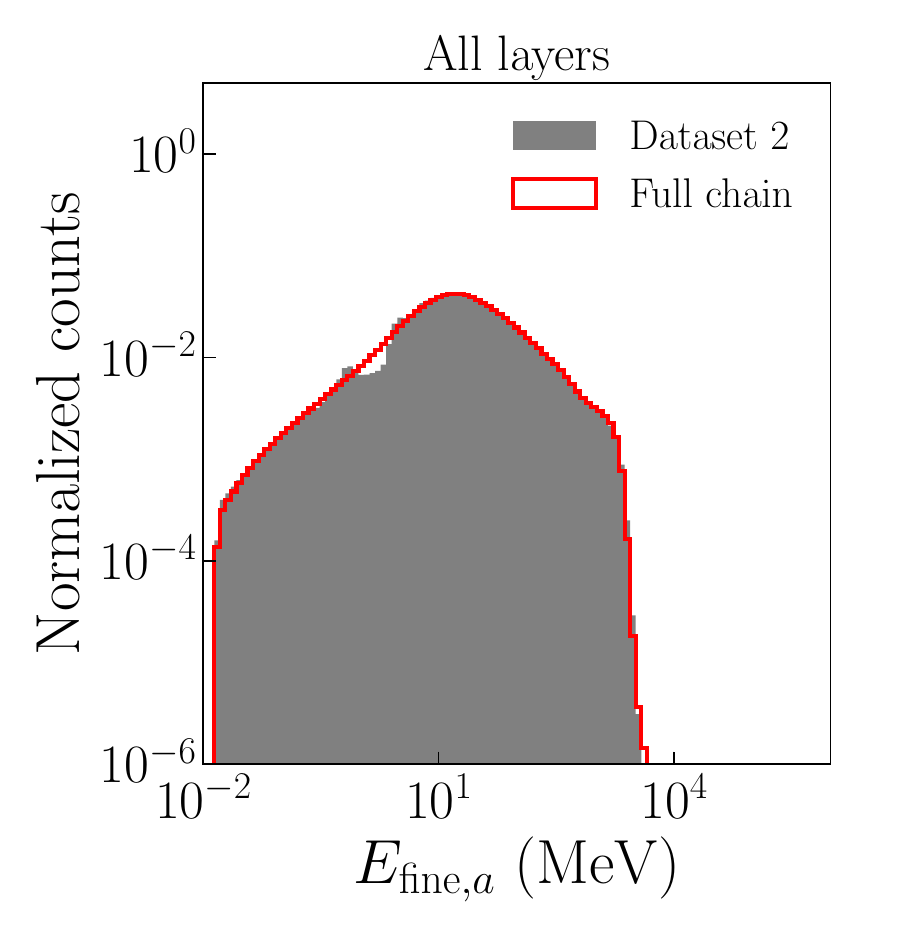}
\caption{Histograms of energy deposition per fine voxel in all layers. 
Distributions of \geant\ data are shown in gray, and those of the full model chain as red lines. \label{fig:voxel_energies}}
\end{figure}

There is good agreement between the generated and reference distributions for the fine voxel energy distributions, as shown in Fig.~\ref{fig:voxel_E}. The most obvious difference is that the generated distributions generally do not capture the small peaks observed in the reference distributions. Fig.~\ref{fig:voxel_energies} shows the energy distribution over all fine voxels in the shower. The generated distribution matches the reference distribution well across six orders of magnitude, with slight differences where there are small bumps in the reference distribution.

\begin{figure*}[ht]
\includegraphics[width=0.5\columnwidth]{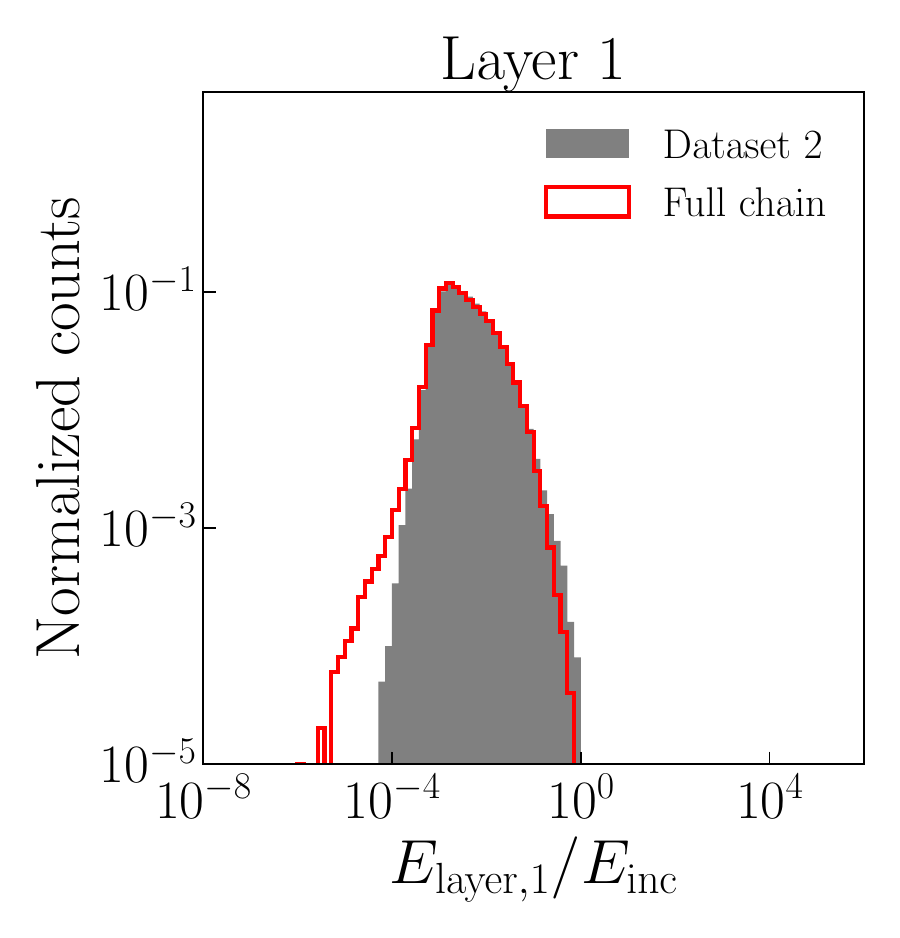}\includegraphics[width=0.5\columnwidth]{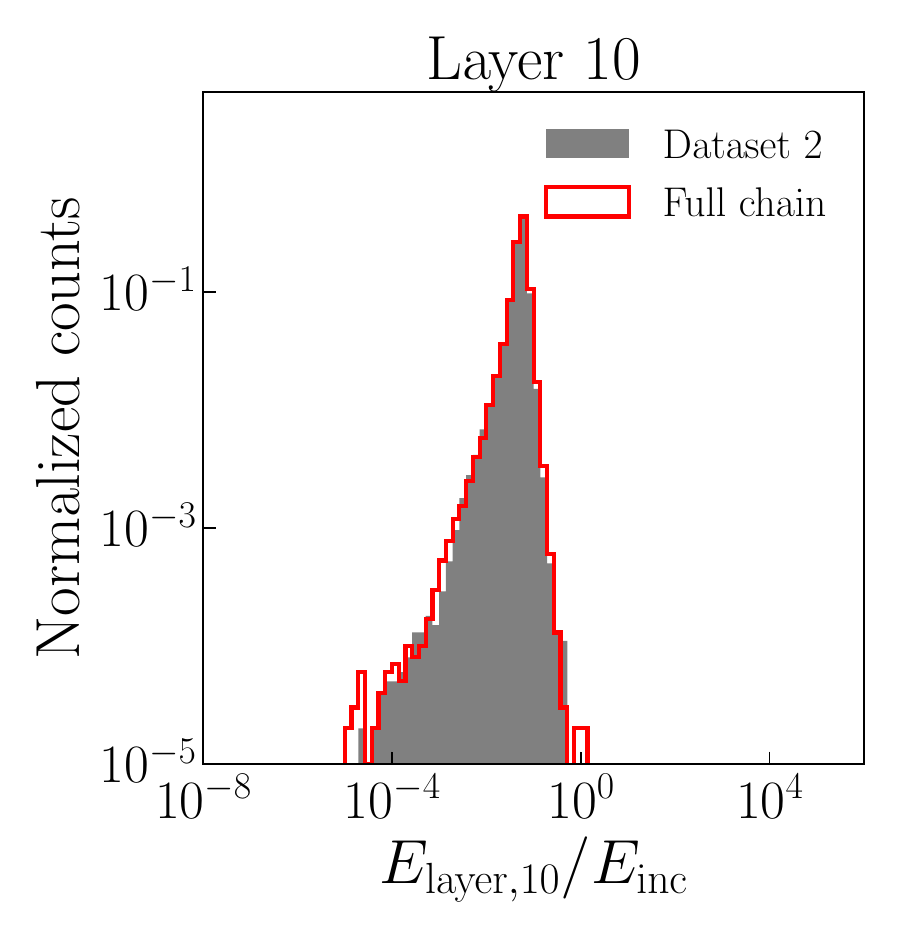}\includegraphics[width=0.5\columnwidth]{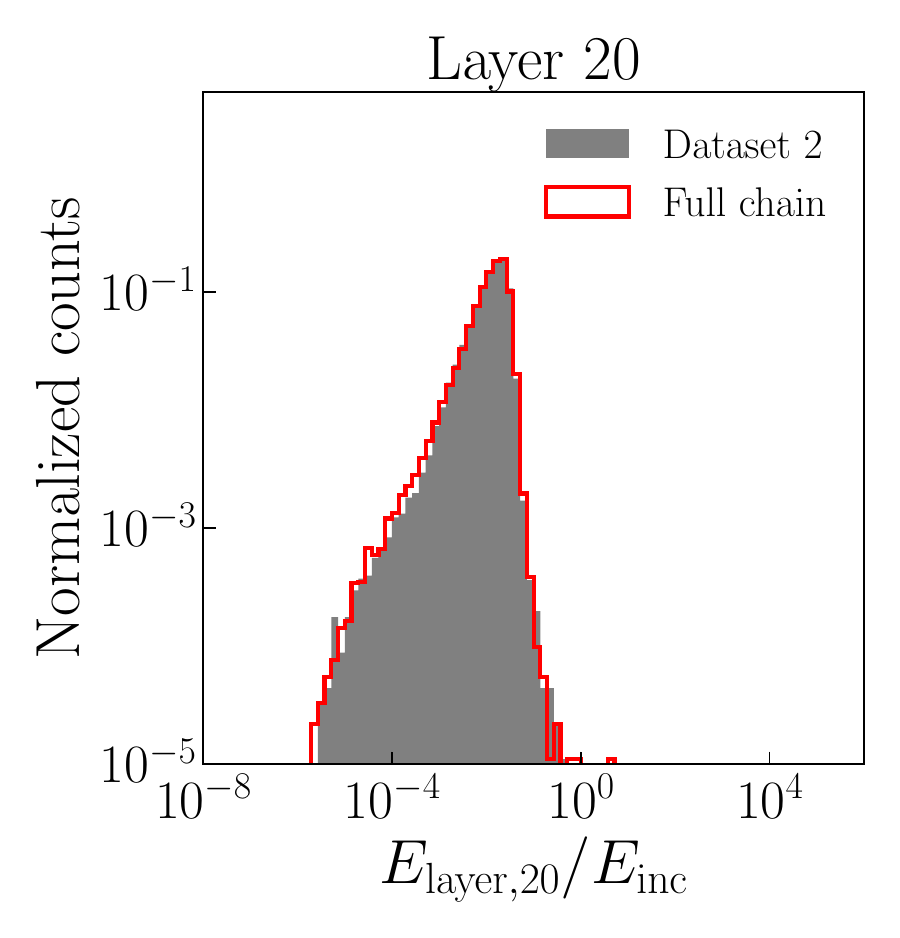}\includegraphics[width=0.5\columnwidth]{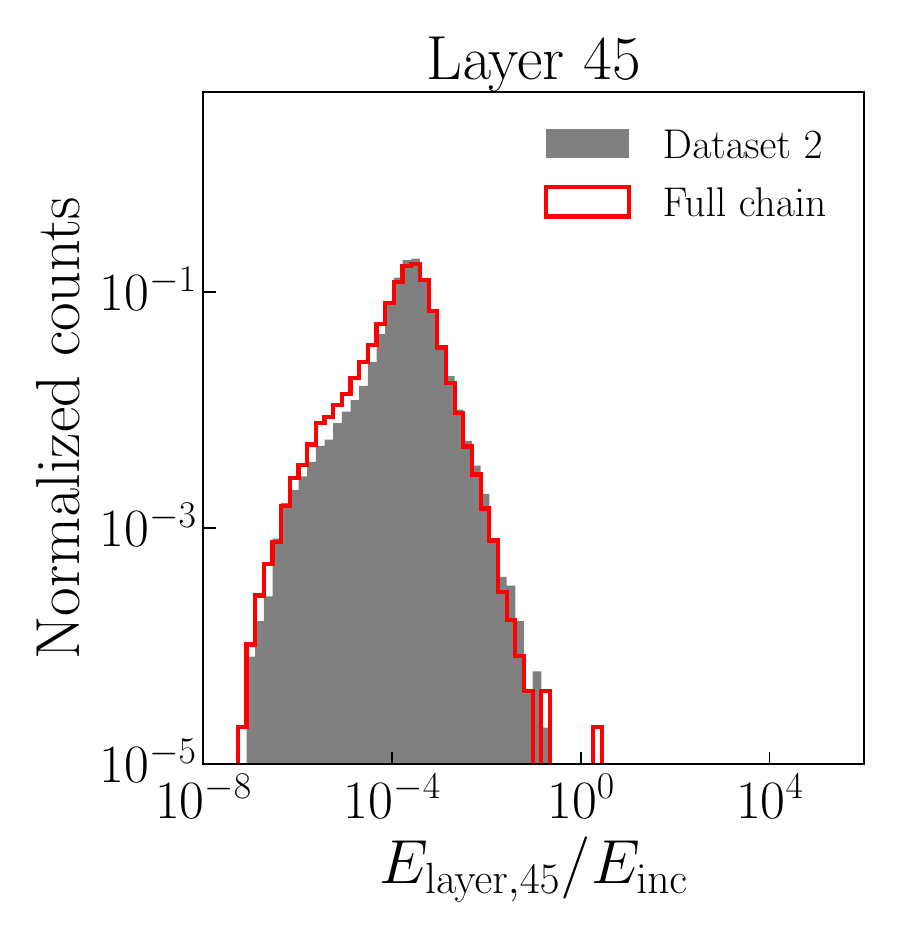}

\caption{Histograms of the ratio of total energy deposited and incident energy in a layer $i$ ($E_{{\rm layer},i}$), for $i=1$, 10, 20, and 45 (from left to right). Distribution of \geant\ data is shown in gray, and that of the full model chain as red lines.
\label{fig:edep_einc_layers}}
\end{figure*}

\begin{figure}[ht]
\includegraphics[width=0.7\columnwidth]{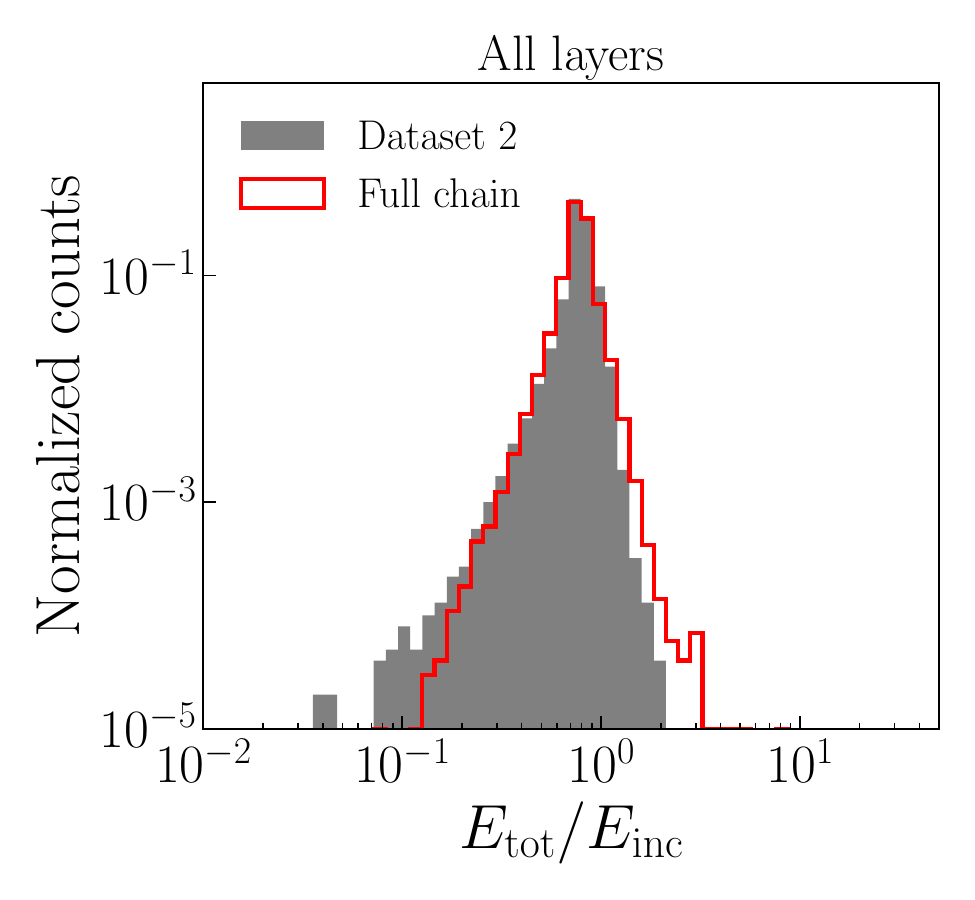}
\caption{Histograms of the ratio of total energy deposition (all layers) and incident energy. 
Distributions of \geant\ data are shown as black lines, and that of the full model chain as red lines. \label{fig:total_edep_einc}}
\end{figure}

Next, we compare the distributions of the ratio of total energy deposited in each of the four selected layers $E_{{\rm layer},i}$ and the incident energy of the particle $E_{\rm inc}$.
These are shown in Fig.~\ref{fig:edep_einc_layers}.
The generated distributions match the reference distributions well for layers 10, 20 and 45. For layer 1, we see that there is an excess at low values that is consistent with the low value excess observed in the layer 1 energy distribution in Fig.~\ref{fig:layer_E}.
In Fig.~\ref{fig:total_edep_einc} we see that there our full model chain is able to generate the correct $E_{\rm tot}/E_{\rm inc}$ distribution, with a slight overestimation at high values and underestimation at low values, though the peak itself is well modelled.

\begin{figure*}[!ht]
\includegraphics[width=0.5\columnwidth]{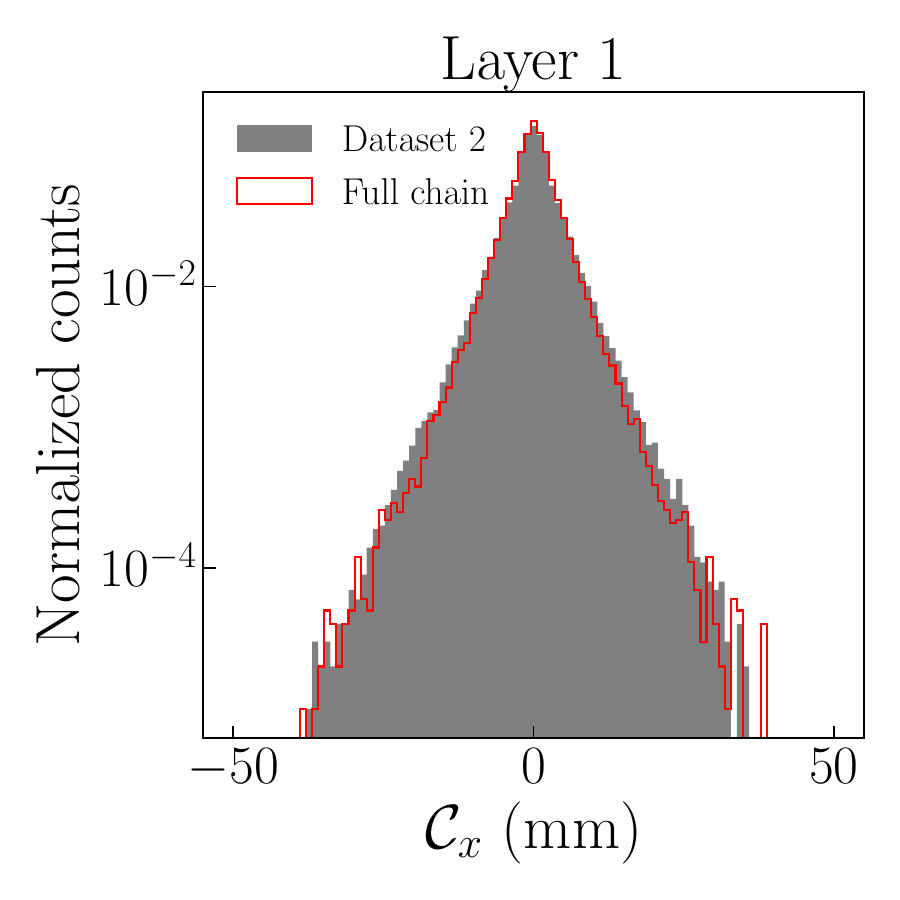}\includegraphics[width=0.5\columnwidth]{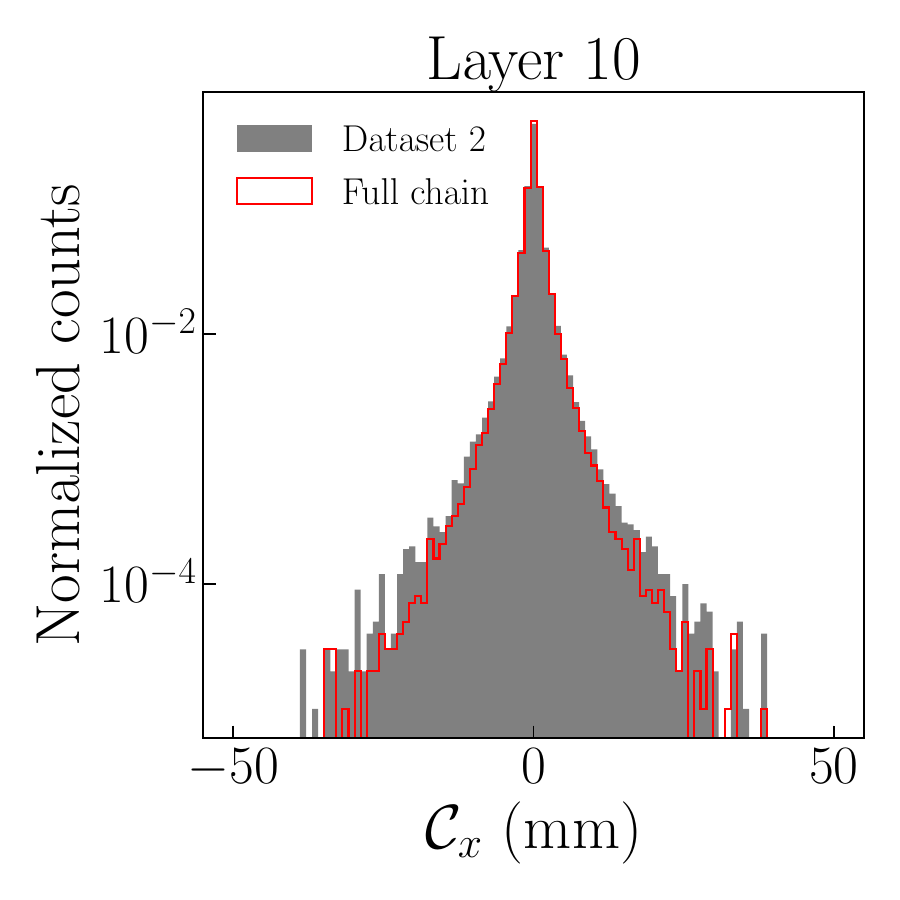}\includegraphics[width=0.5\columnwidth]{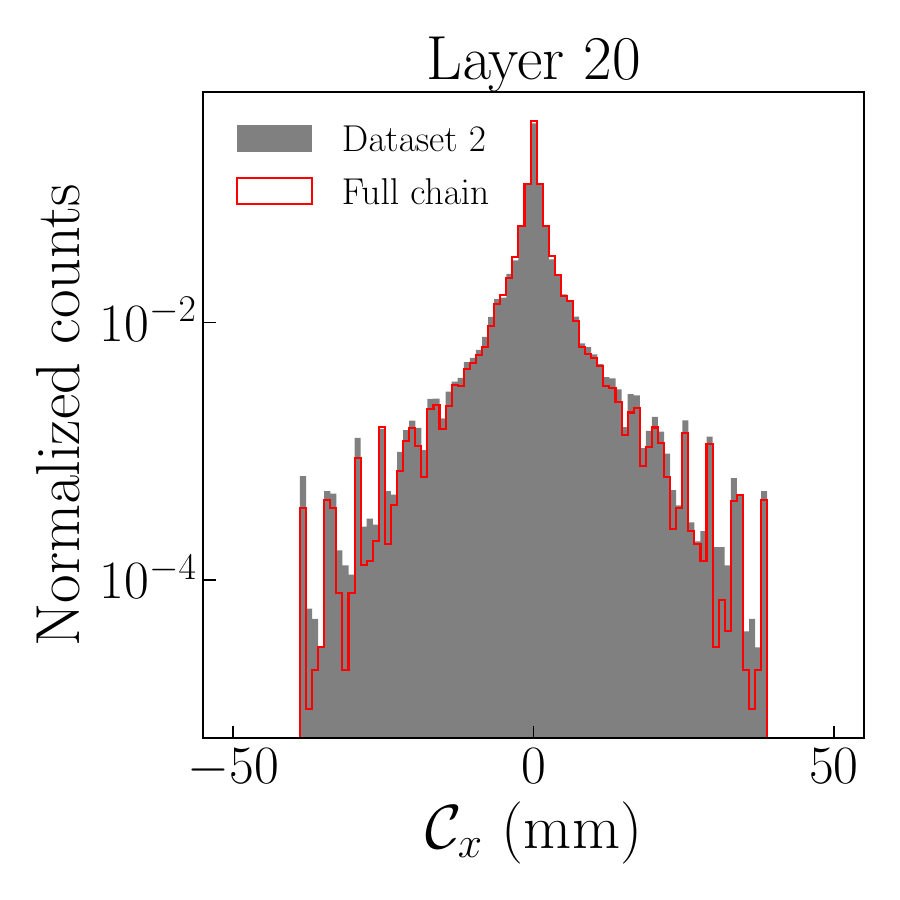}\includegraphics[width=0.5\columnwidth]{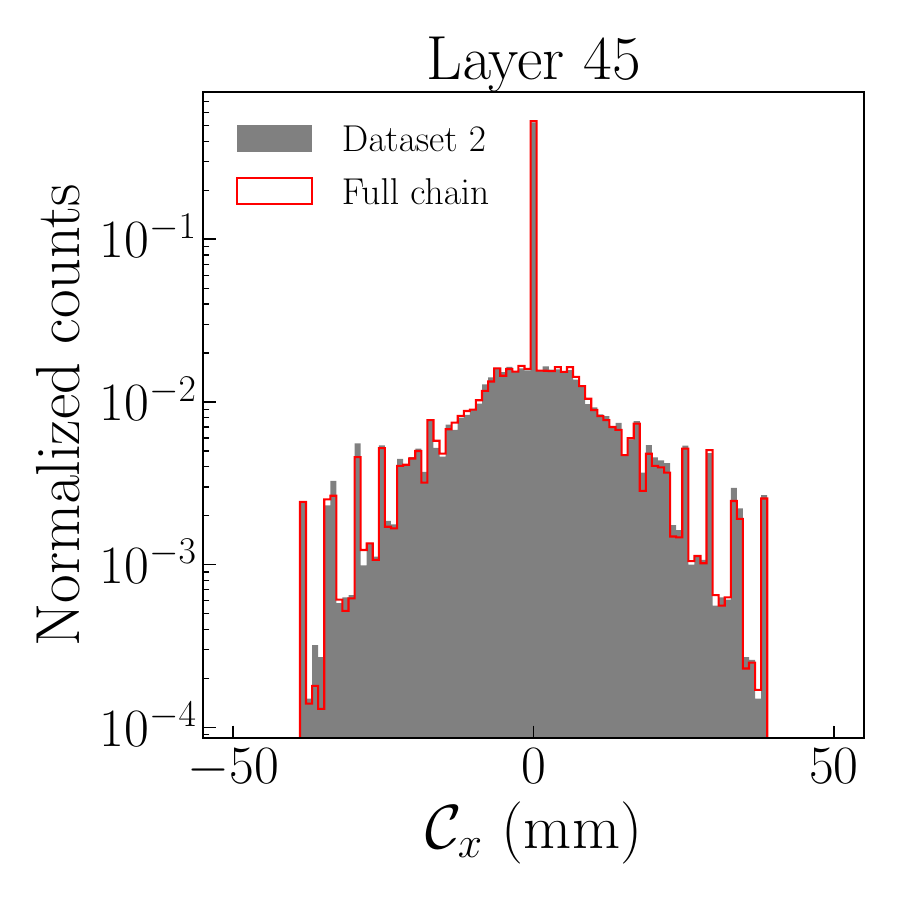}

\caption{Histograms of the centers of energy along the $x$-axis, ${\cal C}_x$. 
Distribution of \geant\ data is shown in gray, and that of the full model chain as red lines. Due to the symmetry of the detector and incident beam, the distributions for the centers of energy in the $y$ direction (${\cal C}_y$) are statistically identical to ${\cal C}_y$. \label{fig:coe}}
\end{figure*}

\begin{figure*}[!ht]
\includegraphics[width=0.7\columnwidth]{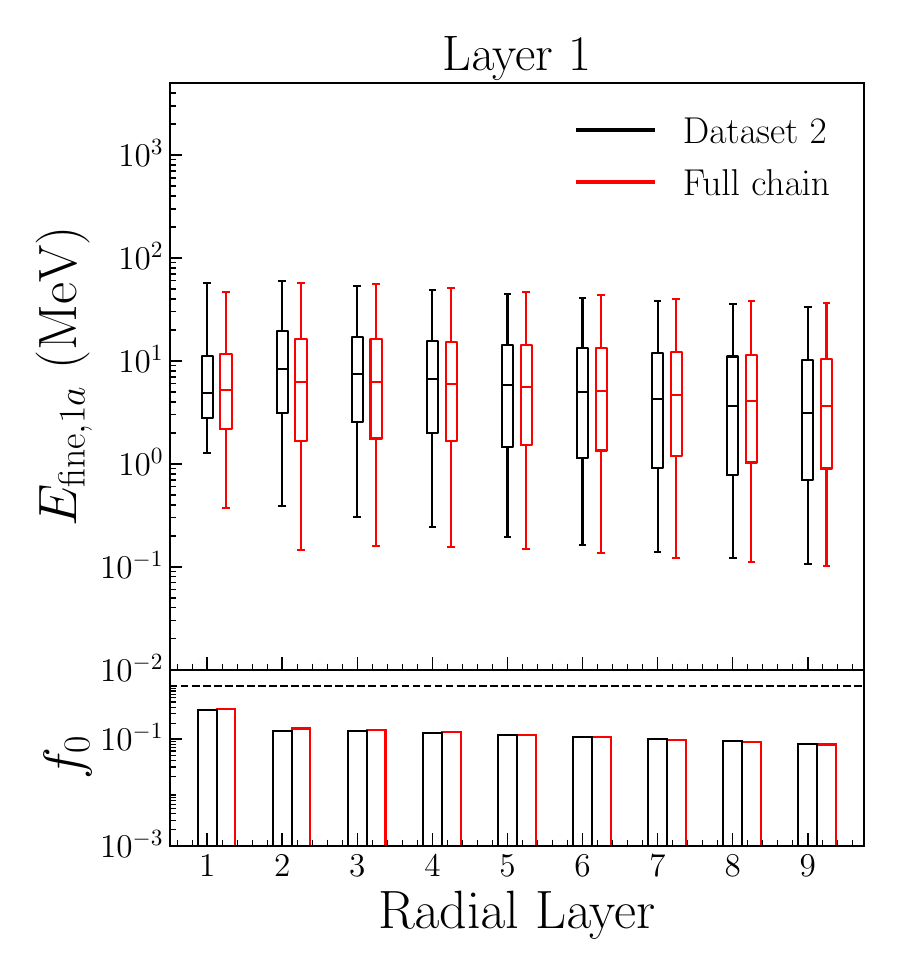}\includegraphics[width=0.7\columnwidth]{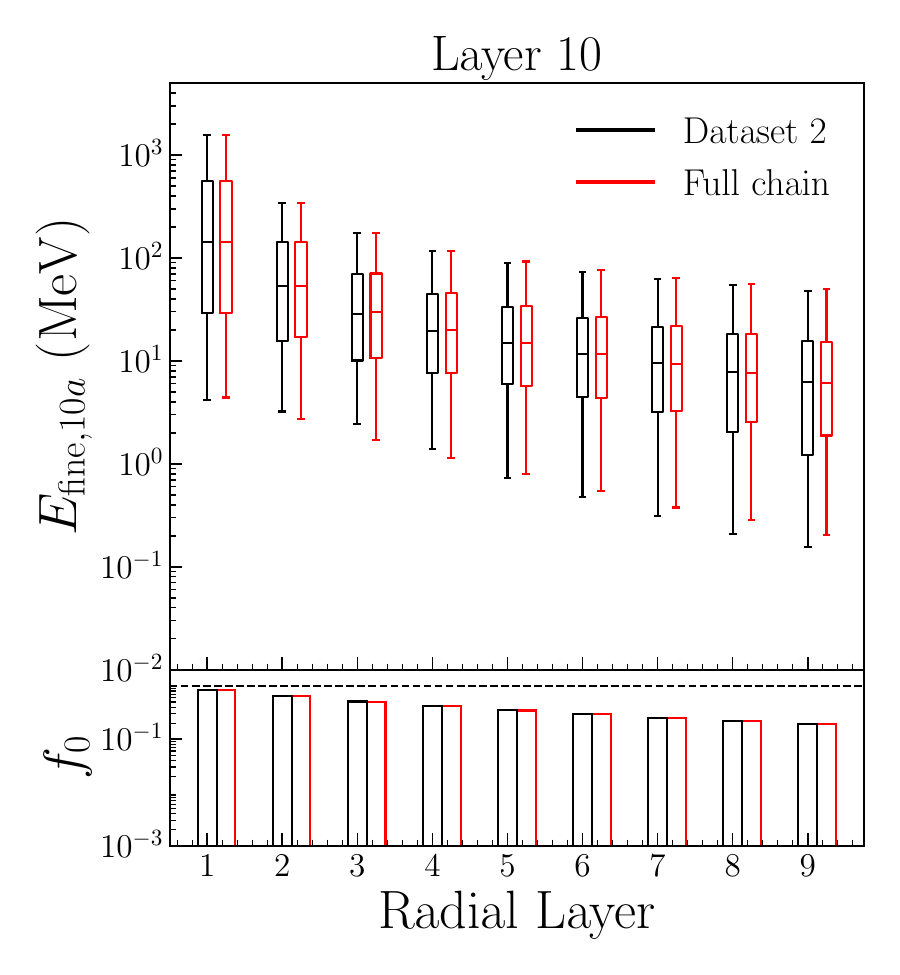}

\includegraphics[width=0.7\columnwidth]{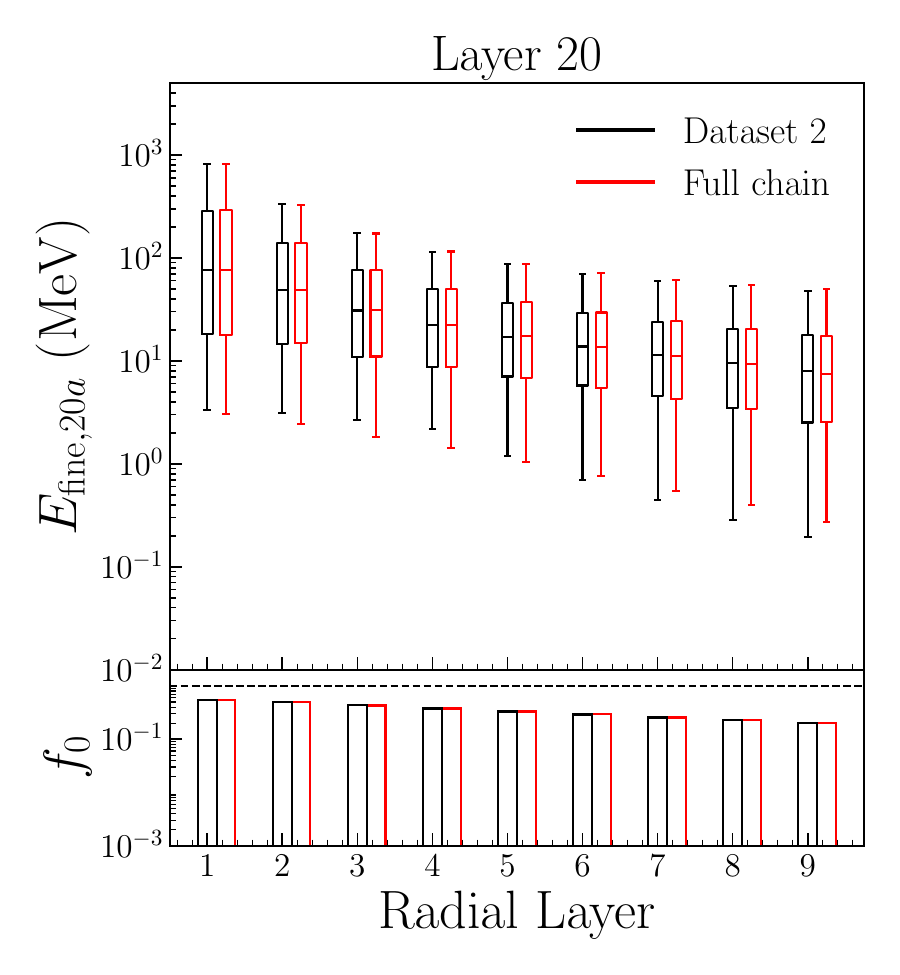}\includegraphics[width=0.7\columnwidth]{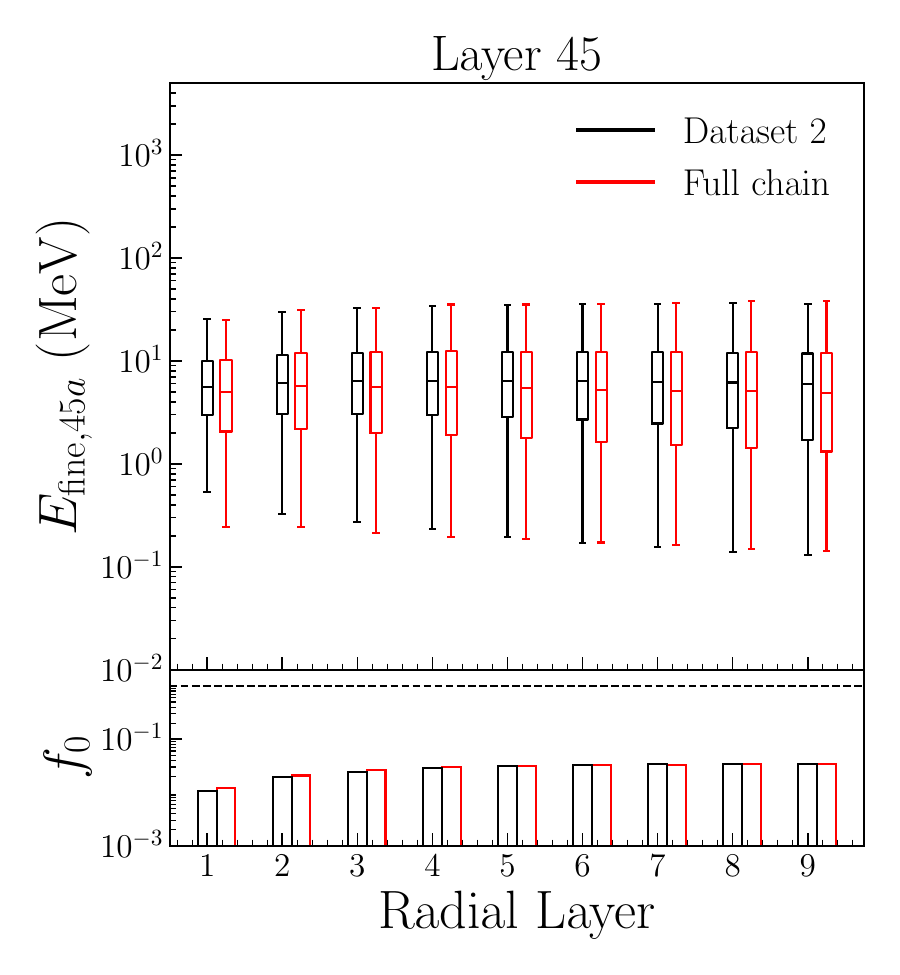}

\caption{Box and whisker plots showing the distribution of energy deposited in each ring of voxels at fixed radial distance from the beam line (nine rings in total) in layers 1, 10, 20, and 45. \geant{} data are shown in black, events generated by the full model chain are shown in red. Each box extends from the first quartile of energies greater than zero to the third quartile. The whiskers extend from the $5^{\rm th}$ to $95^{\rm th}$ percentile of the non-zero energy deposition. Lower subplots show $f_0$, the average fraction of voxels in each radial ring with zero energy deposition.} \label{fig:ds2_boxwhisker}
\end{figure*}

Next we look at the shower shape distribution within each layer. The full model chain is able to accurately generate the distributions of the centers of energy along $x-$axis, $C_x$, as shown in Fig.~\ref{fig:coe}. The same plots for the center of energy along the $y-$axis, $C_y$, are nearly identical due to the rotational symmetry in the $\alpha$ direction.
In Fig.~\ref{fig:ds2_boxwhisker}, we show box-and-whisker plots of voxel energies in each of the nine radial bins within a single layer. We find that the voxel energy distributions can vary significantly across the radial bins, especially in the middle layers (see layers 10 and 20). Nevertheless, we observe good agreement between the generated and reference distributions.

\begin{figure*}[!ht]
\includegraphics[width=0.5\columnwidth]{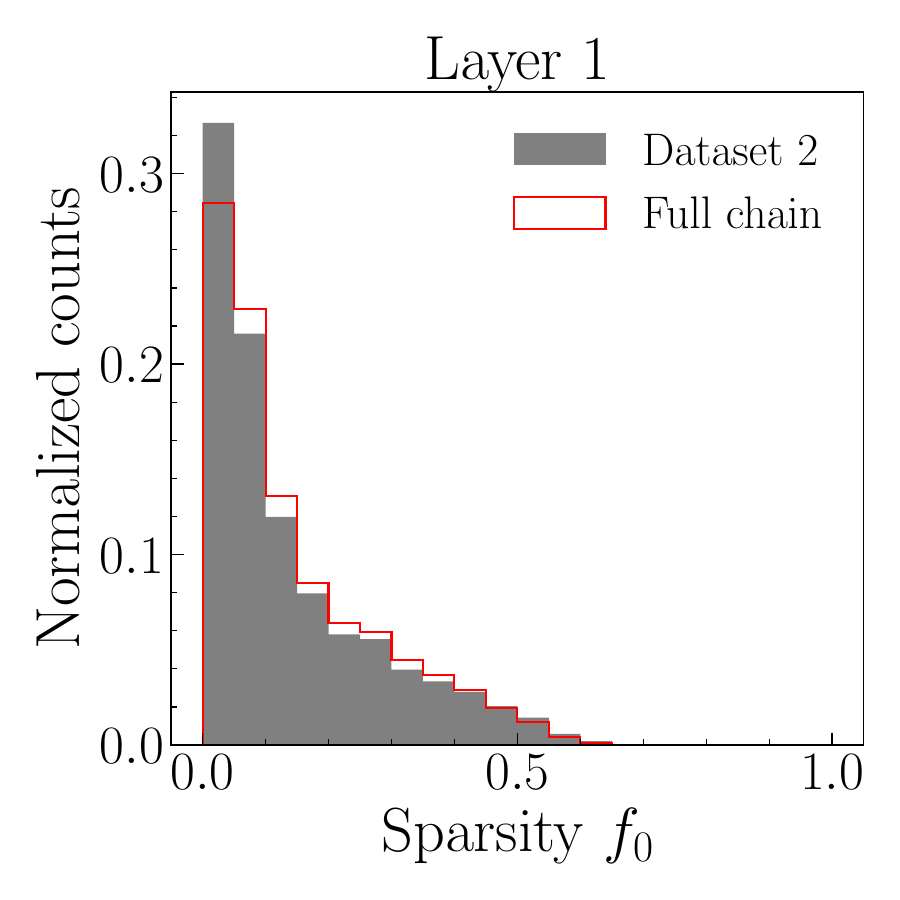}\includegraphics[width=0.5\columnwidth]{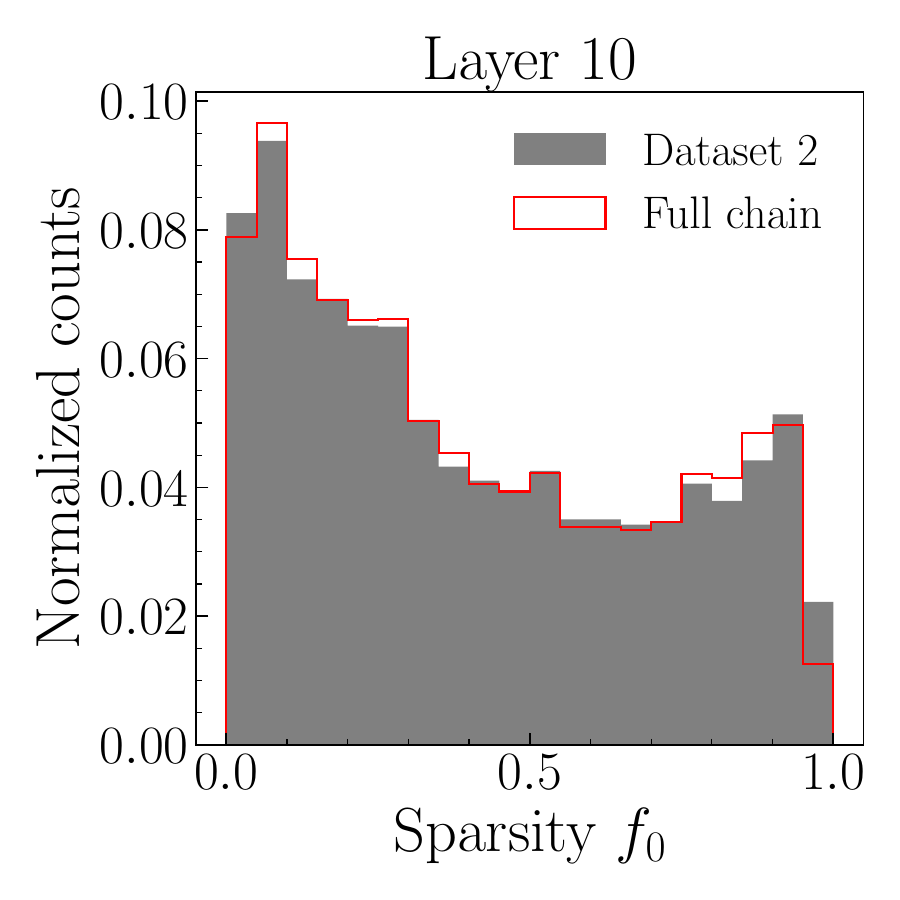}\includegraphics[width=0.5\columnwidth]{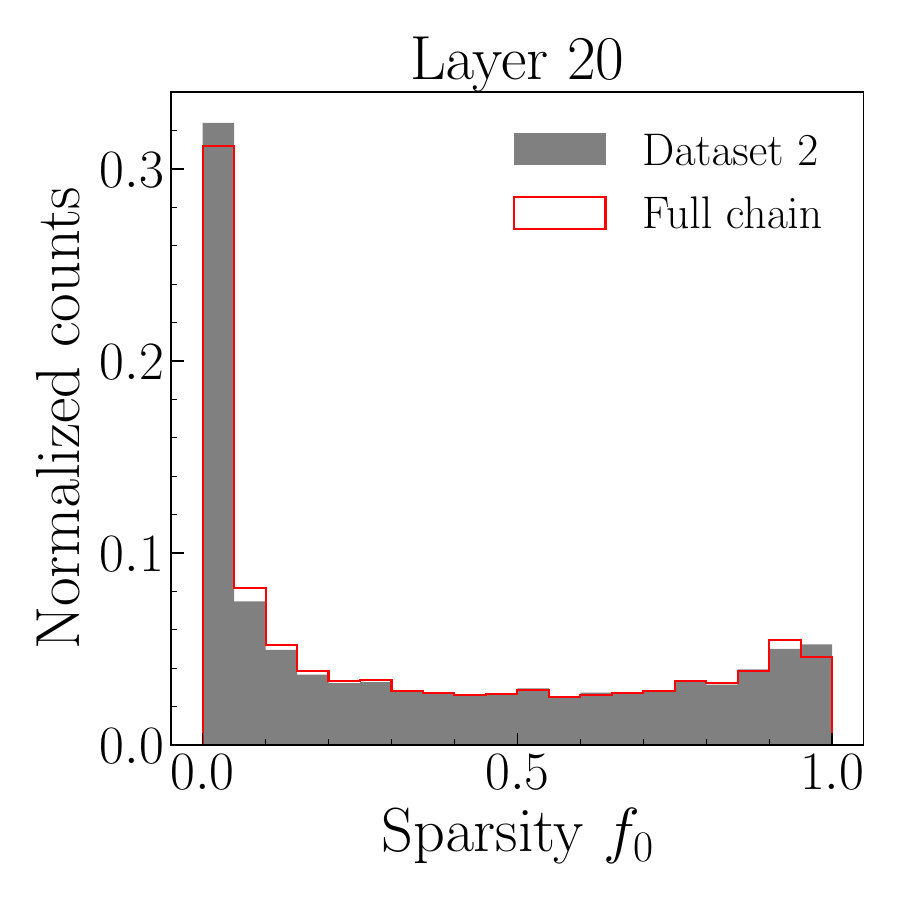}\includegraphics[width=0.5\columnwidth]{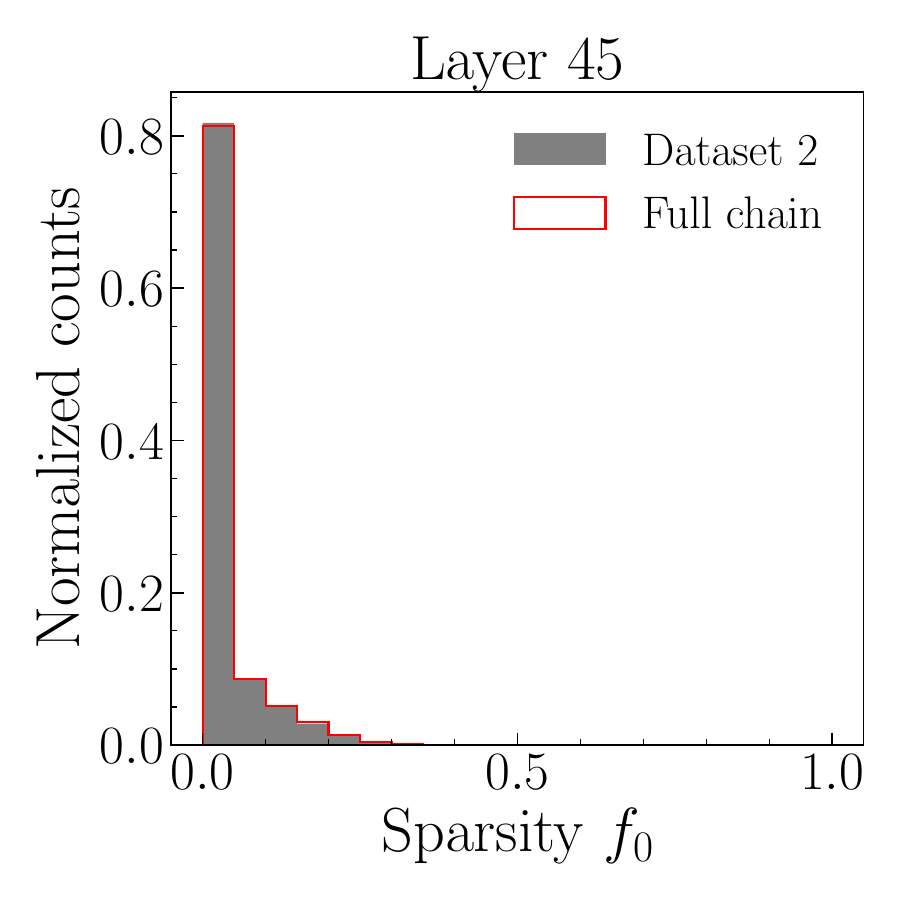}
\caption{Histograms of fraction of voxels in layers 1, 10, 20, and 45 (from left to right) which have non-zero energy deposition $f_0$.
Distributions of \geant\ data are shown in gray, and those of the full model chain as red lines. \label{fig:sparsity}}
\end{figure*}

Lastly, we compare the sparsity of the generated and reference samples in Fig.~\ref{fig:sparsity}. The large fraction of zero voxels ($\sim 75\%$) in Dataset 2 is a major challenge for flow-based models. When training all our flow models in the work, we found that adding noise to voxel (layer) energies during training was necessary to prevent the flow from only learning zero energy voxels (layers). In Fig.~\ref{fig:sparsity}, we see that our full chain approach accurately generates the sparsity within each layer, though with a slight overestimation in layer 1. 

\subsection{Classifier scores}
Similar to Sec.~\ref{sec:compare_coarse}, we train a DNN binary classifier, with the same architecture and training procedure, to separate generated samples from the full chain from reference \geant{} samples.
This allows us to quantify the decrease in sample quality from the combined mismodelling by the three separate flows in the full chain.

\renewcommand{\arraystretch}{1.5}
\begin{table}[!ht]
\begin{center}
\resizebox{\columnwidth}{!}{%
\begin{tabular}{|c||c|c|c|c|}
\hline
&\multicolumn{2}{|c|}{low-level features} &  \multicolumn{2}{c|}{high-level features}\\
 Model & AUC & JSD & AUC & JSD \\
\hline \hline
 Full chain &\; \bf{0.726(19)} \; & \;\bf{0.117(19)} \; & \; \bf{0.715(3)} \; & \; \bf{0.110(4)} \; \\
\hline
 \icalo{} &\; \; 0.797(5) \; & \; 0.210(7) \; & \; 0.798(3) \; & \; 0.214(5) \; \\
\hline
\end{tabular}}
\caption{Mean and standard deviation of 10 independent classifier runs trained on reference \geant{} samples versus generated samples from either the full model chain (this work) or \icalo{} teacher~\cite{Buckley:2023daw}.}
\label{tab:cls_results_full}
\end{center}
\end{table}
\renewcommand{\arraystretch}{1}

From the values in Table~\ref{tab:cls_results_full}, we still find that both LLF and HLF classifier scores are significantly lower than unity. Also, there is only a slight degredation in performance in comparison to the values for \scalo{}~$A$ in Table \ref{tab:cls_results_upsample}. This suggests that the combined mismodelling of \flowone{} and \flowtwo{} with \scalo{} does not significantly reduce the quality of the generated samples. Additionally, the AUCs/JSDs in Table \ref{tab:cls_results_full} are lower than those of the \icalo{} teacher model~\cite{Buckley:2023daw}, which was evaluated on the same classifier architecture.

\subsection{Degree of variation across upsampled showers}
\label{sec:variation}
To test the degree of variation of the fine voxels that are upsampled from coarse voxels, we upsample 100k full-dimensional calorimeter showers from fewer total ($<$ 100k) coarse showers. For example, we could first sample 20k coarse showers using the \flowone{}+\flowtwo{} setup. Next, we upsample each coarse shower five independent times using \scalo{} to generate the full set of 100k fine showers. When upsampled from 20,000 showers, we have LLF and HLF classifier AUCs of 0.762 and 0.724 respectively which are still far from unity. This suggests that these samples are still of sufficiently high fidelity. The classifier scores for full-dimensional calorimeter showers upsampled from different number of coarse showers are shown in Table~\ref{tab:cls_results_coarse}. Here we see that our samples still have high-fidelity even with much fewer coarse showers (e.g.,~$5\times10^3$). We find that the classifier is only able to clearly distinguish the generated samples from the reference samples when we upsample from as few as 1000 coarse showers. This study suggests that \scalo{} is able to generate a substantial amount variation in the upsampled fine voxels.

\renewcommand{\arraystretch}{1.5}
\begin{table}[!ht]
\begin{center}
\resizebox{\columnwidth}{!}{%
\begin{tabular}{|c||c|c|c|c|}
\hline
Number of& \multicolumn{2}{|c|}{low-level features} &  \multicolumn{2}{c|}{high-level features}\\
coarse showers& AUC & JSD & AUC & JSD \\
\hline \hline
$2\times10^4$ & \; 0.762(3) \; & \;0.160(4)  \; & \; 0.724(2) \; & \;0.119(3) \; \\
\hline
$1\times10^4$ & \; 0.795(4) \; & \; 0.208(6) \; & \; 0.738(4) \; & \; 0.135(5) \; \\
\hline
$5\times10^3$ & \; 0.852(4) \; & \; 0.310(6) \; & \; 0.759(3) \; & \;0.162(3) \; \\
\hline
$2\times10^3$ & \; 0.938(2) \; & \; 0.556(7) \; & \; 0.818(3) \; & \; 0.255(6) \; \\
\hline
$1\times 10^3$ & \; 0.980(1) \; & \; 0.769(4) \; & \; 0.887(4) \; & \; 0.408(10) \; \\
\hline
\end{tabular}
}
\caption{Mean and standard deviation of 10 independent classifier runs for showers that were upsampled from fewer than 100k coarse showers.}
\label{tab:cls_results_coarse}
\end{center}
\end{table}
\renewcommand{\arraystretch}{1}

\subsection{Timing}
\label{sec:timing}
Generation speed is a key performance metric for fast calorimeter surrogate models. The \geant{} generation time per event strongly depends on the incident energy of the incoming particle, with the generation time being much longer for the showers with higher incident energies. When averaged over incident energy, it is approximately ${\cal O}(100\,{\rm s})$ per event. In contrast, like most other deep learning-based fast calorimeter simulation models, the generation time of our setup is independent of incident energy. In Table~\ref{tab:generation_time_batches}, we show the generation time per event for the full model chain at different batch sizes on the GPU/CPU. Here the batch size refers to the number of coarse showers generated per batch, where each coarse shower contains 648 coarse voxel energies. Generating the full 100k coarse showers with \flowone{}+\flowtwo{}, we find that the full model setup which includes \scalo{} is able to achieve $\mathcal{O}(10^3)$ speedup compared to \geant{} for a batch size of 1000 events. For the same batch size, we are able to achieve a speed-up factor of $\sim 1.9$ compared to \icalo{} teacher~\cite{Buckley:2023daw}, which also has an MAF-based setup. We estimate that replacing the MAF architecture used here with an IAF or coupling-layer architecture would allow us to achieve a generation time of  $\sim~1.4$ ms per event for a batch size of 1000. This would be a speed-up factor of $\sim~1.6$ compared to \icalo{} student~\cite{Buckley:2023daw} for the same batch size.\footnote{Most of the speed-up comes from replacing the MAF with an IAF in \flowtwo{}. The relative speed-up of \scalo{} itself would be a factor of $\sim10$ compared to the MAF.
Coupling layers are expected to result in similar timings.}

We also include the timings for reduced number of coarse showers (see Section~\ref{sec:variation}) in Table~\ref{tab:generation_time_batches}. As the generation of coarse showers represents the most computationally intensive component of generation, we find that we are able to get a factor of $\sim N$ reduction in total generation time by generating a factor of $N$ fewer coarse showers with the \flowone{}+\flowtwo{} setup and resampling $N$ times from \scalo{}.

The faster generation time together with the variation in upsampled showers suggests that resampling from \scalo{} could be used to reduce the number of coarse-grained simulated showers which need to be generated, saving even more computation time.
This would be particularly beneficial if coarse-grained showers are produced with more time-intensive generative models, such as diffusion.

\renewcommand{\arraystretch}{1.5}
\begin{table}[!ht]
\begin{center}
\resizebox{\columnwidth}{!}{
\begin{tabular}{|c||c|c|c|c|}
\hline
\multirow{2}{*}{Number of}&\multirow{2}{*}{} &  \multirow{2}{*}{Resampling}&\multicolumn{2}{c|}{Generation time per event (ms)}\\
coarse showers &Batch size&factor& GPU& CPU\\
\hline
 \multirow{4}{*}{$1\times10^5$}&1 &\multirow{4}{*}{$\times 1$}& $9.11 \times 10^3$ & $9.73\times 10^4$\\
&10 & & $9.73 \times 10^2$  & $1.11\times 10^4$ \\
&100 & & $1.62 \times 10^2$ &  -\\
&1000 & &$8.75 \times 10^1$  &  - \\
\hline
\multirow{4}{*}{$2\times10^4$}&1 & \multirow{4}{*}{$\times 5 $}& $1.94\times 10^3$ & $2.21\times10^4$\\
&10 &  & $2.08\times10^2$  &  $2.46\times 10^3$\\
&100 & &$4.15\times10^1$  &  -\\
&1000 & &$2.55\times10^1$  &  - \\
\hline
\multirow{4}{*}{$1\times10^4$}&1 & \multirow{4}{*}{$\times 10$} & $1.04\times10^3$ & $1.24\times10^4$ \\
&10 &  & $1.18\times10^2$  & $1.55\times10^3$ \\
&100 &   &$2.63\times10^1$ &  -\\
&1000 & &$1.77\times10^1$ &  - \\
\hline
\end{tabular}}
\caption{Average time taken to generate a single shower event by the full chain for different number of coarse showers generated by the \flowone{}+\flowtwo{} setup. The timing was computed for different generation batch sizes on an Intel i9-7900X CPU at 3.30GHz and a \textsc{TITAN V GPU}. Here the batch size refers to the number of coarse showers generated per batch. For each batch of coarse showers, the number of fine showers generated is equal to the batch size multiplied by the resampling factor.
We were not able to calculate the CPU timings for larger batch sizes due to memory constraints. }
\label{tab:generation_time_batches}
\end{center}
\end{table}
\renewcommand{\arraystretch}{1}

\section{Conclusion}
\label{sec:conclusion}
In this work, we demonstrated the potential of generating high-dimensional calorimeter showers by upsampling coarser showers with \scalo. While we used normalizing flows in this first proof-of-concept work, the \scalo\ approach is completely general and can be applied to any generative modeling framework.
 
For more 
 time-intensive alternative models that are able to generate higher fidelity coarse showers compared to our \flowone{}+\flowtwo{} setup, we can reduce the overall generation time by sampling a factor of $N$ fewer coarse showers and resampling $N$ times from $\scalo$ to obtain fine showers that are still expected to be of very high fidelity. 

As this work only serves as a proof-of-concept for the \scalo\ idea,  we did not extensively optimize the flow architecture and hyperparameters used in this study. An interesting extension to this work would be to implement this setup using an IAF or coupling-layer architecture. This is expected to bring about significant generation speed-up, while possibly preserving the high-fidelity of the upsampled showers.
Attempts to correlate the upsampling across all coarse voxels could also be studied. If successful, this might allow us to use alternative coarse voxelizations (e.g.,~Choice B) that make it easier to accurately model the HLFs.

At the ML4Jets 2023 conference, SuperCalo was presented as one of the best in terms of sample quality for Dataset 2. However, the specific implementation described in this paper suffers in terms of speed compared to the fastest models in the \textit{CaloChallenge} due to the fact that we are using a MAF (instead of IAF). Nevertheless, it is important to note that the SuperCalo framework is implementable with many other architectures and hence the speed may vary based on the architecture choice.

Another promising extension to this work would be to generalize this approach to even higher-dimensional calorimeter showers (e.g.,~Dataset 3 of the \textit{CaloChallenge}~\cite{CaloChallenge_ds3}). This would require a superresolution flow that generates $\mathcal{O}(100)$ fine voxel energies given a single coarse voxel energy.

\section*{Acknowledgements}
IP and DS are supported by DOE grant DOE-SC0010008.
JR is supported with funding through the SNSF Sinergia grant CRSII5\_193716 ``Robust Deep Density Models for High-Energy Particle Physics and Solar Flare Analysis (RODEM)" and the SNSF project grant 200020\_212127 ``At the two upgrade frontiers: machine learning and the ITk Pixel detector".
In this work, we used the {\tt NumPy 1.16.4} \cite{harris2020array}, {\tt Matplotlib 3.1.0} \cite{4160265}, {\tt scikit-learn 0.21.2} \cite{scikit-learn}, {\tt h5py 2.9.0} \cite{hdf5}, {\tt pytorch 1.11.1} \cite{NEURIPS2019_9015}, and {\tt nflows 0.14} \cite{nflows} software packages. Our code is available at {\tt https://github.com/Ian-Pang/supercalo}.

\appendix

\section{Normalizing flows}
\label{sec:NFs}
A normalizing flow (NF)~\cite{tabak_flows,dinh2014nice,razende_flows,dinh2016density,flows_review,CNFs} is a parametric diffeomorphism $f_\theta$ between a latent space, with known distribution $\pi(z)$, and a data space of interest with unknown distribution $p(x)$. 
In the case of a conditional NF, this transformation becomes $f_\theta(x|c)$ which maps to the conditional probability density, where $c$ are the conditional inputs to the flow.

It is defined by a series of invertible functions, parametrized by $\theta$, which can be trained following the change of variables formula
\begin{equation*}
    \log\bigl(p(x|c)\bigr) = \log(\pi(f_\theta(x|c))) + \log\Bigl|\det\left(\mathcal{J}\bigl(f_\theta(x|c)\bigr)\right)\Bigr|,
\end{equation*}
where $\mathcal{J}\bigl(f_\theta(x|c)\bigr)$ is the Jacobian of the transformation $f_\theta(x|c)$.
The allowed transformations must have a tractable Jacobian, which is ideally efficient to compute, and the probability density of the target distribution must be known.
A common choice for $\pi(z)$ is the standard normal distribution.

During generation, it is possible to sample from the known distribution $\pi(z)$ and perform the inverse mapping $f_\theta^{-1}(z|c)$ given some observation $c$ and return a single probabilistic sample under $p(x)$. By sampling the full distribution $\pi(z)$ it is possible to recover the full conditional distribution.

This makes NFs particularly well suited to learning the inverse mappings for which there is not a unique solution, either due to stochasticity or an information bottleneck such as dimension reduction.
As such they are a logical choice for learning $p(\vec E_{\rm fine}|\vec E_{\rm coarse})$.

NFs have already shown promising performance in applications to high energy physics both directly for detector simulation~\cite{Krause:2021ilc,Krause:2021wez,Krause:2022jna,Cresswell:2022tof,Diefenbacher:2023vsw,Buckley:2023daw}, as well as for event generation and analysis  applications~\cite{Bellagente:2020piv,Brehmer:2020vwc,Bothmann:2020ywa,Gao:2020zvv,Gao:2020vdv,Nachman:2020lpy,Choi:2020bnf,Bieringer:2020tnw,Hollingsworth:2021sii,Winterhalder:2021ave,Hallin:2021wme,Jawahar:2021vyu,Butter:2021csz,Winterhalder:2021ngy,Butter:2022lkf,Verheyen:2022tov,Leigh:2022lpn,Kach:2022qnf,Kach:2022uzq,Dolan:2022ikg,Backes:2022sph,Heimel:2022wyj,Algren:2023qnb,Nachman:2023clf,Raine:2023fko}.

\section{Pre- and postprocessing}
\label{sec:preprocessing}
The incident energy of the incoming particle $E_{\rm inc}$ is a conditional input for all the NFs used in this work and is preprocessed as 
\begin{equation}
    E_{\rm inc} \to \log_{10} \frac{ E_{\rm inc}}{10^{4.5}~{\rm MeV}} \; \in [-1.5, 1.5]
    \label{eq:Einc.pre}
\end{equation}

In \flowone{}, the total energy deposited in layer $i$ ($E_{{\rm layer},i}$) is calculated by summing the energy values of all voxels in that layer. To address difficulties in learning distributions with many exactly zero elements, uniform noise within the range of $[0,5]$ keV is added to $E_{{\rm layer},i}$~\cite{Krause:2021ilc}. This noisy energy value, $x_i$, is then normalized by dividing it by $65$ GeV, which is slightly larger than the maximum energy deposition observed in any layer of any event in the dataset:

\begin{equation}
E_{{\rm layer},i} \to x_i \equiv (E_{{\rm layer},i} + {\rm rand}[0,5~{\rm keV}])/65~\text{GeV} \label{eq:layer_norm}
\end{equation}

In the final step, a logit transformation is applied to $x_i$ to obtain $y_i$:

\begin{equation}
y_i = \log \frac{u_i}{1-u_i},\quad u_i\equiv \alpha+(1-2\alpha)x_i, \label{eq:logit}
\end{equation}

where the offset $\alpha \equiv 10^{-6}$ is introduced to ensure that the boundaries $x_i = 0$ and $1$ map to finite numbers.

For \flowtwo{}, we obtain the coarse voxel energies by summing up all the fine voxels associated to each coarse voxel. The coarse voxel energies are then preprocessed as

\begin{equation}
E_{{\rm coarse},i} \to \log_{10}\left((E_{{\rm coarse},i} + {\rm rand}[0,5~{\rm keV}])/E_{\rm coarse,max}\right) +6
\end{equation}
where $E_{\rm coarse,max}$ is the maximum coarse voxel energy in the training dataset.

The coarse layer energies $\vec E^{(\rm coarse)}_{\text{layer}}$ are obtained by summing the coarse voxel energies (with added noise) in each coarse layer. The coarse layer energies are then processed as 

\begin{equation}
E^{(\rm coarse)}_{\text{layer}, i} \to \frac{1}{4}\log_{10}\left(E^{(\rm coarse)}_{\text{layer}, i} \right)
\end{equation}

For \scalo{}, we preprocess the coarse voxel energies $E_{{\rm coarse},i}$ according to the following transformations:

\begin{align}
E_{{\rm coarse},i} &\to \tilde{x}_i \equiv (E_{{\rm coarse},i} + {\rm rand}[0,1~{\rm keV}])/E_{\rm coarse,max}\\
\tilde{y}_i &= \log \frac{\tilde{u}_i}{1-\tilde{u}_i},\quad \tilde{u}_i\equiv \alpha+(1-2\alpha)\tilde{x}_i
\end{align}
The energies of neighboring coarse voxels are preprocessed in the same way but without the addition of noise.

Like in \flowone{}, the fine layer energies $E_{{\rm layer},i}$ are preprocessed according to Eqns.~\ref{eq:Einc.pre} and \ref{eq:layer_norm} but without the addition of noise. The coarse layer and coarse r-bin labels used in the conditional inputs of \scalo{} are one-hot encoded.

Lastly, the fine voxel energies $e_{{\rm fine},ij}$ associated to the $i$th coarse voxel is preprocessed as 
\begin{align}
e_{{\rm fine},ij} &\to \hat{x}_{ij} \equiv (e_{{\rm fine},ij} + {\rm rand}[0,0.1~{\rm keV}])/E_{\rm coarse,i}\\
\hat{y}_{ij} &= \log \frac{\hat{u}_{ij}}{1-\hat{u}_{ij}},\quad \hat{u}_{ij}\equiv \alpha+(1-2\alpha)\hat{x}_{ij}
\end{align}

During generation, the preprocessing is inverted for the outputs of the flow to recover the correct physical quantities. Since the minimum fine voxel energy in Dataset 2 is 15 keV, a cut is applied to the generated outputs (in physical units) of \flowone{}, \flowtwo{} and \scalo{} such that energies below 15 keV are set to zero.
\section{Classifier architecture}
\label{sec:classifier_arch}

The classifier performance evaluation in this study relies on a neural network architecture obtained from the {\it CaloChallenge}~\cite{calochallenge} evaluation script.

To elaborate, the classifier is a deep, fully-connected neural network comprising an input layer and two hidden layers, each containing 2048 nodes. All activation functions in the hidden layers are leaky ReLUs, with a default negative slope of 0.01. However, the output layer uses a sigmoid activation function to produce a single output number. No regularization techniques such as batch normalization or dropout are applied.

The input data for low- and high-level feature classification is preprocessed following the methodology described in the main text. A dataset consisting of 100k showers~\cite{CaloChallenge_ds2} is used, which is split into train, test, and validation sets in a 60:20:20 ratio. This is an independent dataset that was not used to train the models in this work.

The neural networks are optimized through training for 50 epochs using the \textsc{Adam} optimizer~\cite{kingma2014adam} with an initial learning rate of $2\times 10^{-4}$. The batch size of 1000 is used and the binary cross-entropy is minimized during training.

For the final evaluation, the model state with the highest accuracy on the validation set is selected, and the classifier is subsequently calibrated using isotonic regression~\cite{2017arXiv170604599G} from the {\tt sklearn} library~\cite{scikit-learn} based on the validation dataset before being evaluated on the test set.

\bibliographystyle{JHEP}
\bibliography{literature}
\end{document}